\documentclass{article}
\usepackage[english]{babel}
\usepackage[utf8]{inputenc}
\usepackage{amsmath}
\usepackage{algorithm}
\usepackage{algpseudocode}
\usepackage{graphicx}
\usepackage{johd}
\usepackage{amssymb}
\usepackage{amsfonts}
\usepackage{verbatim}
\usepackage{bm}
\usepackage{calrsfs}

\DeclareMathAlphabet{\pazocal}{OMS}{zplm}{m}{n}

\DeclareMathOperator*{\argmin}{arg\,min}

\title{Spectral estimation for mixed causal-noncausal autoregressive models}

\author{Alain Hecq \footnote{a.hecq@maastrichtuniversity.nl}$\;$ and Daniel Velasquez-Gaviria\footnote{Corresponding author: d.velasquezgaviria@maastrichtuniversity.nl School of Business and Economics, Tongersestraat 53, 6211 LM Maastricht, Netherlands}\\
        \small Maastricht University\\
}
\date{\today}

\begin{document}
\maketitle

\begin{abstract} 

\noindent 
  This paper investigates new ways of estimating and identifying causal, noncausal, and mixed causal-noncausal autoregressive models driven by a non-Gaussian error sequence. We do not assume any parametric distribution function for the innovations. Instead, we use the information of higher-order cumulants, combining the spectrum and the bispectrum in a minimum distance estimation. We show how to circumvent the nonlinearity of the parameters and the multimodality in the noncausal and mixed models by selecting the appropriate initial values in the estimation. In addition, we propose a method of identification using a simple comparison criterion based on the global minimum of the estimation function. By means of a Monte Carlo study, we find unbiased estimated parameters and a correct identification as the data depart from normality. We propose an empirical application on eight monthly commodity prices, finding noncausal and mixed causal-noncausal dynamics.
\end{abstract}

\noindent\\

\section{Introduction}

The traditional assumption of causality in autoregressive models is widely adopted in classical estimation methods for economic time series. The causality implies that the time series depends on a combination of lags with roots outside the unit circle, forcing the time series to depend only on past information. Despite this, causality is not always justified and is not systematically tested. Additionally, it emerges that including only past information for estimating economic models fails to explain their structural shocks, as shown by \cite{hansen2019two}. In addition, \cite{lanne2012optimal,lanne2012has}  show that including future information may improve the forecast accuracy, i.e., through noncausal model representations. A statistically rigorous presentation of noncausal models arises in \cite{breid1991maximum}, dividing the autoregressive polynomial into two components, one causal with roots outside the unit circle and another noncausal with roots inside the unit circle, or equivalently, including leads in the autoregressive process. Furthermore, allowing the autoregressive polynomial to contain roots simultaneously inside and outside the unit circle, introduces the mixed causal-noncausal models. The noncausal dynamics offer the possibility to capture nonlinearities in weakly stationary variables, such as locally explosive patterns and bubbles, replicating characteristics of anticipation and speculation. Consequently, noncausal models can produce valuable risk measures, fueling their interest in economics, with applications in inflation, commodity prices, and stock indices, see among others \cite{gourieroux2017local,lof2013noncausality,lof2017noncausality,lanne2011noncausal,hecq2016identification, hecq2021forecasting}. \\

An interesting aspect of the noncausal models is the estimation and identification. The reason is that they share the same second-order dependence structure, the autocovariance function or its frequency domain analog, the spectral density, with the causal models. Consequently, the estimations based only on second-order moments such as Gaussian-likelihood, OLS, Yule-Walker, or Whittle yield coefficients corresponding to roots outside the unit circle identifying a causal model, even though the actual roots of the coefficients are inside the unit circle, corresponding to a noncausal model. The underlying reason for this lack of identification is the assumption of Gaussianity. For the Gaussian series, the probability structure is entirely determined by second-order moments. In general, for an AR($p$) model with a Gaussian errors sequence, there exist $2^{p}$ possible specifications of its roots with the same probability structure. In contrast, when the error sequence is non-Gaussian, each of the $2^{p}$ configurations of roots corresponds to a different weakly stationary process with a unique and identifiable probability structure. Evidently, knowing the non-Gaussian probability distribution of the error sequence, the natural estimator is the Maximum Likelihood (ML), identifying the true roots of the autoregressive polynomial regardless of whether it is causal or noncausal, see \cite{basawa1976asymptotic}. This estimation was initially proposed by \cite{lii1992approximate} for possible noninvertible MA models, proving that it is consistent for any symmetric, strictly positive, and sufficiently smooth non-Gaussian probability density function. \cite{huang2000quasi} propose a quasi-Likelihood estimation of possible noninvertible MA models using the Laplacian distribution. \cite{breid1991maximum} proposes an Maximum likelihood estimation (MLE) for possibly noncausal autoregressive processes driven by non-Gaussian i.i.d. noise, evidencing that the estimated parameters are asymptotically normal and providing identification when the error sequence is heavy-tailed. \cite{lii1996maximum, lehr1998maximum}, implement MLE for possible noncausal and noninertible ARMA models. Similarly, \cite{lanne2011noncausal}, \cite{hecq2016identification}, \cite{giancaterini2022inference} propose a parameter estimation in possible noncausal AR models using the Student's t distribution. For the same distribution, \cite{andrews2006maximum} proposes the estimation of the all-pass ARMA models, where all roots of the autoregressive polynomial are reciprocals of roots of the moving average polynomial and vice versa. MLE for autoregressive models using the alpha-stable distribution has also been implemented in \cite{andrews2009maximum, fries2021conditional}.\\

A natural criticism of MLE is the prior knowledge about the probability distribution of the error sequence, which is uncommon empirically. Indeed, arbitrarily assuming a distribution function that does not replicate the features of the data can lead to misestimating parameters and misidentifying the autoregressive model. Alternatively, in this research, we propose to estimate the parameters of the causal, noncausal, and mixed autoregressive model without prior knowledge about the probability distribution of the error sequence. Instead, it has been proposed to employ the higher-order cumulants necessary for constructing the higher-order spectral densities, with the ability to discriminate between causal and noncausal models. Higher-order cumulants, also named higher-order correlations, have been used to estimate noncausal AR models, see \cite{rosenblatt1980linear,lii1982deconvolution,giannakis1990estimating, tugnait1986identification, terdik1999bilinear,nikias1987non}. We use in this paper the procedure proposed by \cite{brillinger1985fourier, leonenko1998spectral,velasco2018frequency}, constructing a function of the second and third-order spectral densities (the spectrum and the bispectrum, respectively) in a minimum distance estimation with the periodogram and biperiodogram. The estimation of the parameters is performed by minimizing the function, achieving the identification of the model in its global minimum.\\

Although this estimation is consistent and asymptotically normal, the global minimum is not guaranteed to be obtained. On the contrary, the estimation function is multimodal (not convex), leading to multiple local minima that increase as the number of parameters in the model grows. Accordingly, the solution is not unique and convergent. For a discussion of the multimodality in noncausal models, see \cite{kindop2021ubiquitous}, and for mixed causal-noncausal, see \cite{bec2020mixed}. Consequently, an empirical prerequisite for reaching the global minimum in the estimation function is the appropriate selection of the initial values. In this sense, our first contribution to the literature is to develop a general scheme for selecting the initial values for estimating noncausal and mixed causal-noncausal based on their causal representations. This allows us to find the correct combination of parameters in the region where the global minimum of the estimation function lies. In our second contribution, we present a decision method for identifying the model based on the existence of the global minimum in the estimation function. We explore the robustness of our method in finite samples through an extensive Monte Carlo study using the alpha-stable distribution. We find unbiased estimated parameters and a correct identification as the data depart from normality. Our third contribution is to perform the third-order spectral estimation of mixed models. To the best of our knowledge, this is the first time in mixed models. In addition, we propose an empirical application of eight monthly commodity prices, evidencing noncausal and mixed causal/noncausal dynamics. \\

The paper is divided as follows: Section 2 presents the methodology, Section 3 presents the estimation method, Section 4 introduces the Monte Carlo study, Section 5 presents the empirical application, and the last Section wraps up the conclusions and discussion. 

\section{Methodology}

Let $y_t$ be a time series of length $T$, with zero mean and finite $k$-th order moments, for $k\geq2$, and $k$-th weakly stationary for $t=\left\{0,\pm 1, \pm2, \pm 3, ... \right\}$, with autocovariance function 

\begin{equation}
    \kappa_2(j)=\mathbb{E}\left[y_t y_{t-j}\right]\;\; for\;\;\ j=\left\{0,\pm 1, \pm2, \pm 3, ... \right\}.
\end{equation}

\begin{center}
     \begin{figure}[h]
         \centering         \includegraphics[width=0.8 \textwidth]{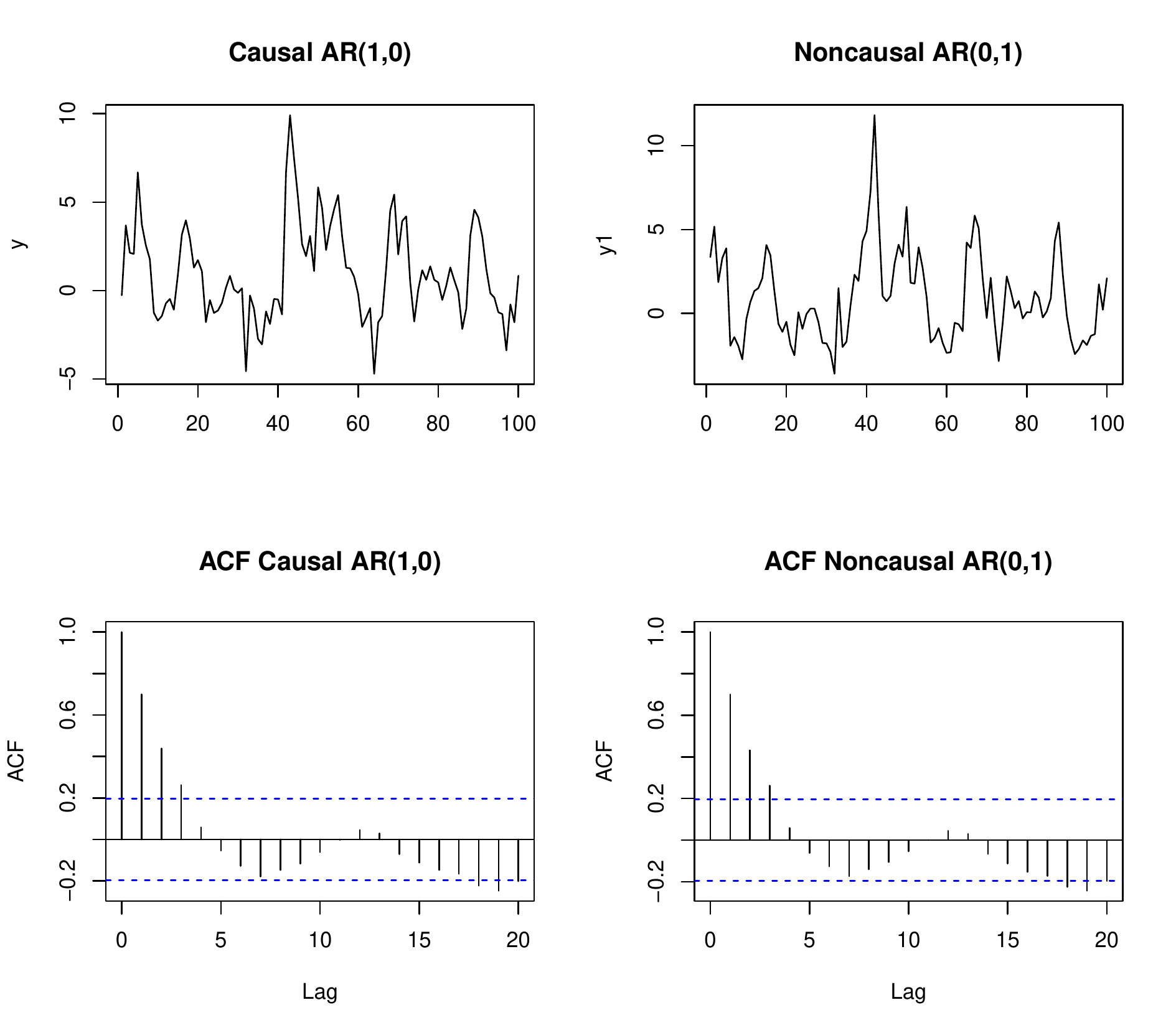}
         \caption{\footnotesize In the left is the causal AR(1,0) model $y_t=0.7y_{t-1}+\varepsilon_t$, and in the right, the noncausal AR(0,1) model $y_t=0.7y_{t+1}+\epsilon_t$, plot and ACF. $\varepsilon_t\sim$ alpha-stable $f_a(t;\alpha=1.5,\beta=0.25,\gamma=1,\delta=0)$. The definition of the alpha-stable distribution is in Section \ref{MC}, the sample size $T=100$. The advantage of using the alpha-stable distribution is the ability to have different shapes depending on the value of its parameters. It can easily replicate sample characteristics of leptokurticity and skewness. Although theoretically, the existence of its moments is linked to the values of its stability parameter ($\alpha$), i.e., the variance, the skewness, and the kurtosis only exist when it is equal to two. An interesting feature of the alpha-stable distribution is its infinite divisibility, i.e., it can be expressed as a sum of i.i.d variables while preserving its distribution. See Section \ref{MC} for details.}
        \label{causal-noncausal-acf}
     \end{figure}
 \end{center}

Notice that the autocovariance is considering linear relations of the data in two time periods, known as second-order dependence, also that $\kappa_{2_j}=\kappa_{2_{j-1}}...=\kappa_{2_{j-n}}=\kappa_{2_{j+1}}...=\kappa_{2_{j+n}}$. In any of these representations, the autocovariance function remains unchanged. Therefore, estimations based on the second-order structure, e.g., Gaussian-Likelihood, OLS, or Yule-Walker, fail to distinguish between causal and noncausal models. The above is evident in Figure \ref{causal-noncausal-acf}. On the left is a causal model and on the right is a noncausal model with the same coefficients and the same alpha-stable error sequence. Notice that the autocorrelation functions are identical.

However, when the time series is non-Gaussian, the probability structure of causal and noncausal models is different and identifiable using additional information from the higher-order cumulants. We specify that the first three moments are equivalent to the first three cumulants. This is not valid for cumulants of orders greater than three. In particular, the third cumulant contains the relation of the time series at three points in time. Hence it is known as the triple correlation function or the bicovariance function, see \cite{bartelt1984phase}. The third-order cumulant is

\begin{equation}
    \kappa_3(j,l)=\mathbb{E}\left[y_t y_{t-j}y_{t-l}\right]\;\; for\;\;\ j,l=\left\{0,\pm 1, \pm2, \pm 3, ... \right\}.
\end{equation}

Note that $\kappa_3(j,l)$ is equal to zero for the Gaussian and symmetrical distributions. The definition of $\kappa_3(j,l)$ is the third coefficient of the power expansion of the natural logarithm of the characteristic function of a random variable. Given the stationarity of $y_t$, $\kappa_3(j,l)$ satisfies the following symmetry conditions

\begin{equation} \label{cumu}
 \kappa_3(j,l) = \kappa_3(l,j)=\kappa_3(-j,l-j)=\kappa_3(l-j,-j)=\kappa_3(j-l,-l)=\kappa_3(-l,j-l).
\end{equation}

The second and third-order cumulants are necessary for constructing the spectrum and the bispectrum. The spectrum is the Fourier transform of the second cumulant. Analogously, the bispectrum is the Fourier transform of the third-order cumulant. Defining the Fourier frequencies as $\omega=2\pi/T$, the spectrum is

\begin{equation}
    S_2(\omega_j)=\frac{1}{2\pi}\sum_{j=-\infty}^{\infty} \kappa_2(j) e^{-ij\omega},
\end{equation}

where $i$ is the imaginary number. The bispectrum is

\begin{equation}
    S_3(\omega_1,\omega_2)=\frac{1}{(2\pi)^2}\sum_{j=-\infty}^{\infty}\sum_{l=-\infty}^{\infty} \kappa_3(j,l) e^{-i(j\omega_1+l\omega_2)}.
\end{equation}

Note that for the existence of the bispectrum, the third-order cumulant must be different from zero. We assume that the spectrum and the bispectrum have continuous second-derivatives and are bounded above and away from zero in modulus for the sets of frequencies $\Lambda_1$ and $\Lambda_2$. Where $\omega_1 \subset \Lambda_1=(0,1/2)$ for $S_2$,  and given the symmetries $S_3(\omega_1,\omega_2)=S_3(\omega_2,\omega_1)=S_3(\omega_1,-\omega_2-\omega_1)=S_3(-\omega_1-\omega_2,\omega_2)$ and the identity $S_3(\omega_1,\omega_2)=S_3(-\omega_1,-\omega_2)$, $\Lambda_2$ is a finite union of compact sets inside an open triangle with vertices $(0,0),(1/2,0),(1/3,1/3)$, for details see \cite{rao1989estimation,rosenblatt1965estimation,rao2012introduction, anh2007minimum}. 

The nonparametric estimation of the spectrum and bispectrum for autoregressive processes is usually based on methods employing the discrete Fourier transform (DFT), i.e., assuming that the time series comprises a set of harmonically related sinusoids. Let us define the DFT as

\begin{equation}\label{dft1}
    d_T(\omega)=\sum_{t=0}^{T-1}y_te^{-it\omega}.
\end{equation}

The most popular estimator of the spectrum is the periodogram. For a summary of spectrum estimates, see \cite{kay1981spectrum}. The periodogram is obtained as the modulus of the output values from the DFT in Equation \ref{dft1},
performed directly on the time series. The origin of the periodogram can be traced back to \cite{schuster1898investigation}, which conducted a Fourier series fitting to the variation of sunspot numbers trying to find hidden periodicities in the data. Thereafter, it has been widely used in spectral analysis and autoregressive modeling, see \cite{brillinger1975time, rosenblatt1965stationary,alekseev1996asymptotic,bartlett1950periodogram,brillinger1967asymptotic}. The periodogram is defined as

\begin{equation}
    I_2(\omega)=\frac{1}{2\pi T} d_T(\omega)\overline{d_T(\omega)}.
\end{equation}

Similar to the periodogram, however, taking into account relations between three frequencies (bifrequencies), the biperiodogram is the bispectrum estimator, defined as

\begin{equation}
    I_3=(\omega_1,\omega_2)=\frac{1}{(2\pi)^2 T}d_T(\omega_1)d_T(\omega_2)\overline{d_T(-\omega_1-\omega_2)},
\end{equation}

where $\overline{d_T(-\omega_1-\omega_2)}$, represents the complex conjugate of the Fourier transform of the sum of two frequencies. This component is the one that accounts for the asymmetries present in non-Gaussian data. Contrary to the periodogram, which is always real, the biperiodogram is complex due to its last factor. The periodogram and biperiodogram are asymptotically unbiased estimators of the spectrum and bispectrum but inconsistent; i.e., the variance of the estimates will not tend to zero as $T$ increases without bound, see \cite{brillinger1975time, rosenblatt1965stationary,alekseev1996asymptotic,bartlett1950periodogram}. Some properties of the periodogram are developed in \cite{brillinger1967asymptotic}; in particular, the mean of the periodogram $k=2$ and biperiodogram $k=3$. Denoting $\boldsymbol{\omega}={\omega_1,\omega_2}$

\begin{equation}  E[I_k(\boldsymbol{\omega})]=S_k(\boldsymbol{\omega})+o(1),
\end{equation}

when $\omega_s\neq 0\;mod\;2\pi$. The variance for $k={2,3}$

\begin{equation}
    T^{2-k}Var[I_k(\boldsymbol{\omega})]=
    (k-1)!S_2(\omega_1)...S_2(\omega_{k-1})S_2(\omega_1+...+\omega_{k-1})+o(1),
\end{equation}

as $T\rightarrow\infty$. Equivalently to the periodogram, the biperiodogram at different Fourier frequencies $(\omega\neq\omega')$ is uncorrelated

\begin{equation}
    Cov[I_k(\omega),I_{\iota}(\omega')]=o(T^{\frac{k+\iota-4}{2}}), \;\;\;k,\iota=2,3.
\end{equation}

\subsection{The causal AR($r,0$)}
In this Section, we present causal, noncausal, and mixed autoregressive models. Let $y_t$ be a causal autoregressive process generated by

\begin{equation}
    \phi(L)y_t=\varepsilon_t.
\end{equation}
 The lag polynomial is $\phi(L)=1-\phi_1 L-\phi_2 L^2...\phi_r L^r$, with a set of parameters $\Phi_r=(\phi_1,\phi_2...\phi_r)$, where $L$ stands for the backward shift operator, that is, $L^k=y_{t-k}$ for $k=\left\{0,\pm 1, \pm2, \pm 3, ... \right\}$. This model is well known in the time series literature as the AR($p$). In our notation, the AR($p$) model is replaced by the AR($r,0$) model. The main difference between the AR($p$) model and the AR($r,0$) model is that in the former, it is sufficient to assume white noise in the error sequence, while in the latter, we assume that the error sequence is i.i.d and has finite moments greater than 3, that is
 \begin{equation}
     \mathbb{E}(\varepsilon_t)=0,\;\; \mathbb{E}(\varepsilon_t^2)=k_{2_e},\;\; \textrm{and}\;\; \mathbb{E}(\varepsilon_t^3)=k_{3_e}.
 \end{equation}
 
The same assumption applies to the AR($0,s$) and MAR($r,s$) models, discussed in Sections \ref{ars} and \ref{marrs}, respectively. Note that we differentiate the cumulants of $y_t$, $\kappa_2,\kappa_3$, with the cumulants of the error sequence $\varepsilon_t$, $\kappa_{2_e},\kappa_{3_e}$. $\phi(z)$ is provided by the roots $1-\phi_1 z-\phi_2 z^2...-\phi_r z^r$, all lying outside the unit circle. Hence, the model has a covariance-stationary representation form $\kappa_2(j)=\mathbb{E}\left[y_t y_{t-j}\right]$, and is represented as an infinite sum of past shocks, MA($\infty$) as

\begin{equation}\label{eq2}
    y_t=\Psi(L)\varepsilon_t,
\end{equation}

where $\sum_{j=0}^\infty\lvert\Psi \rvert<\infty$. Then, the autocovariance generating function of $y_t$ is $g_y(z)=\sum_{j=-\infty}^{\infty} \kappa_2(j) z^j$. Evaluating $g_y$ at $z=e^{- i\omega}$, and dividing by $2\pi$ we obtain the spectrum of $y_t$

\begin{equation}
    S_2(\Phi_r,\omega)=\frac{1}{2\pi}g_y(e^{-i\omega})=\sum_{j=-\infty}^{\infty}\kappa_2(j) e^{-i\omega j}.
\end{equation}

This function is known as the second-order spectrum since it only considers second-order relations. The spectrum is continuous, non-negative, real-valued, symmetric, and periodic around $\omega$. For the AR($r,0$) model, the spectrum can be obtained explicitly using the representation of Equation \ref{eq2}. Thus, we can define the transfer function of the linear filter AR($r,0$) as 

\begin{equation}
    \psi(\Phi_r,\omega)=1/(1-\phi_1z-\phi_2z^2...-\phi_rz^r).
\end{equation}

Then, the spectrum of the AR($r,0$) is the multiplication of the second-order cumulant of the errors sequence and the transfer function, and its conjugate
\begin{equation} \label{ed_arp}
\begin{split}
 S_2(\Phi_r,\omega) & = \frac{k_{2_e}}{2\pi}\psi(\Phi_r,\omega)\overline{\psi(\Phi_r,\omega)} \\
 & = \frac{k_{2_e}}{2\pi}\frac{1}{(1-\phi_1z-\phi_2z^2...-\phi_rz^r)(1-\phi_1z^{-1}-\phi_2z^{-2}...-\phi_rz^{-r})}\\
 & = \frac{k_{2_e}}{2\pi}\frac{1}{\lvert 1-\phi_1z-\phi_2z^2-...-\phi_rz^r\rvert^2}.
\end{split}
\end{equation}

Defining $\overline{z}=e^{2\pi i (-\omega_1-\omega_2)}$, the transfer function of the sum of the complex conjugation of the frequencies is $\overline{\psi(\Phi_r,-\omega_1-\omega_2)}=1/(1-\phi_1\overline{z}-...-\phi_2\overline{z}^2-...-\overline{z}^r)$. In this way, the bispectrum is

\begin{equation} \label{bispectrum1}
\begin{split}
S_3(\Phi_r,\omega_1,\omega_2) & = \frac{k_{3_e}}{(2\pi)^2}\psi(\Phi_r,\omega_1)\psi(\Phi_r,\omega_2)\overline{\psi(\Phi_r,-\omega_1-\omega_2)} \\
 & = \frac{k_{3_e}}{(2\pi)^2}\frac{1}{(1-\phi_1z-\phi_2z^2...-\phi_rz^r)(1-\phi_1z^{-1}-\phi_2z^{-2}...-\phi_rz^{-r})(1-\phi_1\overline{z}-\phi_2\overline{z}^2-...-\phi_r\overline{z}^r)}.\\
\end{split}
\end{equation}

\subsection{The noncausal AR($0,s$)}
\label{ars}
Let $y_t$ be a noncausal autoregressive process generated by

\begin{equation}
\label{noncausal-arp}
    \varphi(L^{-1})y_t=\varepsilon_t,
\end{equation}

where the lead polynomial is $\varphi(L^{-1})=1-\varphi_1L^{-1}-\varphi_2L^{-2}...\varphi_s L^{-s}$ with the set of parameters $\Phi_s=(\varphi_1,\varphi_2...\varphi_s)$. In this notation, the AR($0,s$) model has $s$ noncausal and zero causal components. The polynomial $\varphi(z)$ is $1-\varphi_1 z-\varphi_2z^{2}...-\varphi_s z^{s}$, with all roots outside the unit circle. Note that the polynomial of the noncausal model can be represented with its roots outside the unit circle, thus depending on its leads, or with roots inside the unit circle depending on its lags. Indeed, the polynomial $\varphi(z)$ has a causal representation with its zeros inside the unit circle, (see \cite{brockwell2009time} p. 81 for a stationary solution), that is $\varphi^{*}(z)=1-\varphi_1^{*}z-\varphi_2^{*}z^2...\varphi_s^{*}z^s$, with $\varphi^{*}(z)\neq0$ for $\lvert z\rvert>1$. Following Equation 6 in \cite{lanne2011noncausal}, the polynomial $\varphi^{*}(z)$ can be expressed as

\begin{equation}\label{eq6}
    \varphi^{*}(z)=-\varphi^{*}_sz^s\left(1+\frac{\varphi^{*}_{s-1}}{\varphi^*_s}z^{-1}+...+\frac{\varphi^{*}_{1}}{\varphi^*_s}z^{1-s}-\frac{1}{\varphi^{*}_s}\right)=-\varphi^{*}_sz^s \varphi(z^{-1}),
\end{equation}

where $\varphi(z^{-1})$ is as in Equation \ref{noncausal-arp}, implying that $\varphi^{*}_{s-j}/\varphi^{*}_s=-\varphi_j$ for $j=1,2,...s-1$, and $1/\varphi^{*}_s=\varphi_j$. Notice that the zeros of $\varphi^{*}(z)$ lie inside the unit circle. In this way, the noncausal model can be rewritten to obtain the following causal specification

\begin{equation}
\label{caus}
    \varphi^{*}(L)y_t=\xi_t,
\end{equation}
where the error term $\xi_t=-1/\varphi_{s}\varepsilon_{t-s}$, and coefficients for the polynomial $\varphi^{*}_j$ equal to $\varphi^{*}_j=-\varphi_j/\varphi_s$ for $j=1,2...s-1$, and $\varphi^{*}_s=1/\varphi_s$. Note that this causal representation with roots inside the unit circle should not be confused with the explosive model, where the error term is contemporaneously uncorrelated with lags of $y_t$ (see \cite{brockwell2009time}, p. 81). As an example, in the \textbf{Remark 1}, we rewrite the noncausal AR(0,$s$) into a purely causal AR($r,0$), for $s=r$.\\

\textbf{Remark 1.} The causal AR($r,0$) representation of the noncausal AR($0,s$) model, for $s=r$. The noncausal AR($0,s$) is
\begin{equation*}
y_t=\varphi_1y_{t+1}+\varphi_2y_{t+2}+...+\varphi_sy_{t+s}+\varepsilon_t. 
\end{equation*}

Reverting the process in $s$ time periods

\begin{equation*}
    y_{t-s}= \varphi_1y_{t+1-s}+\varphi_2y_{t+2-s}...\varphi_sy_{t}+\varepsilon_{t-s}.
\end{equation*}

Solving for $y_t$, we can rewrite $y_t$ as

\begin{equation}
y_t= -\frac{\varphi_1}{\varphi_s}y_{t+1-s}-\frac{\varphi_2}{\varphi_s}y_{t+2-s}+...+\frac{1}{\varphi_s}y_{t-s}-\frac{1}{\varphi_s}\varepsilon_{t-s},
\end{equation}

which highlight the correlation between the error term and the regressors. Then, the purely causal AR($r,0$) representation of the AR($0,s$) is

\begin{equation} \label{eq9}
 y_t = \varphi^{*}_1 y_{t+1-s}+\varphi^{*}_2 y_{t+2-s}+...+\varphi^{*}_s y_{t-s}+\xi_t.
\end{equation}
Notice that the coefficients and the error sequence from Equation \ref{eq9} are equivalent to the ones in Equation \ref{caus}. The causal representation is key to the estimation and identification of the model, as we shall discuss in Section \ref{est_stra}.\\

The transfer function for the noncausal AR(0,$s$) is $\psi(\Phi_s,\omega)=1/(1-\varphi_1z-\varphi_2z^{2}...-\varphi_rz^{s})$, and
its spectrum is
\begin{equation} \label{ed_arp_1}
\begin{split}
 S_2(\Phi_s,\omega) & = \frac{k_{2_e}}{2\pi}(\Phi_s,\omega)\overline{\psi(\Phi_s,\omega)} \\
 & = \frac{k_{2_e}}{2\pi}\frac{1}{(1-\varphi_1z-\varphi_2z^2...-\varphi_sz^s)(1-\varphi_1z^{-1}-\varphi_2z^{-2}...-\varphi_sz^{-s})}\\
 & = \frac{k_{2_e}}{2\pi}\frac{1}{\lvert 1-\varphi_1z^{-1}-\varphi_2z^{-2}...-\varphi_sz^{-s}\rvert^2}.
\end{split}
\end{equation}

Additionally, the bispectrum is

\begin{equation} \label{bispectrum2}
\begin{split}
  S_3(\Phi_s,\omega_1,\omega_2) & = \frac{k_{3_e}}{(2\pi)^2}\psi(\Phi_s,\omega_1)\psi(\Phi_s,\omega_2)\overline{\psi(\Phi_s,\omega_1+\omega_2)} \\
 & = \frac{k_{3_e}}{(2\pi)^2}\frac{1}{(1-\varphi_1z^{-1}-\varphi_2z^{-2}...-\varphi_sz^{-s})(1-\varphi_1z-\varphi_2z^2...-\varphi_sz^s)(1-\varphi_1\overline{z}-\varphi_2\overline{z}^2-...-\varphi_s\overline{z}^s)}.\\
\end{split}
\end{equation}

\begin{center}
     \begin{figure}[h]
         \centering         \includegraphics[width=0.7 \textwidth]{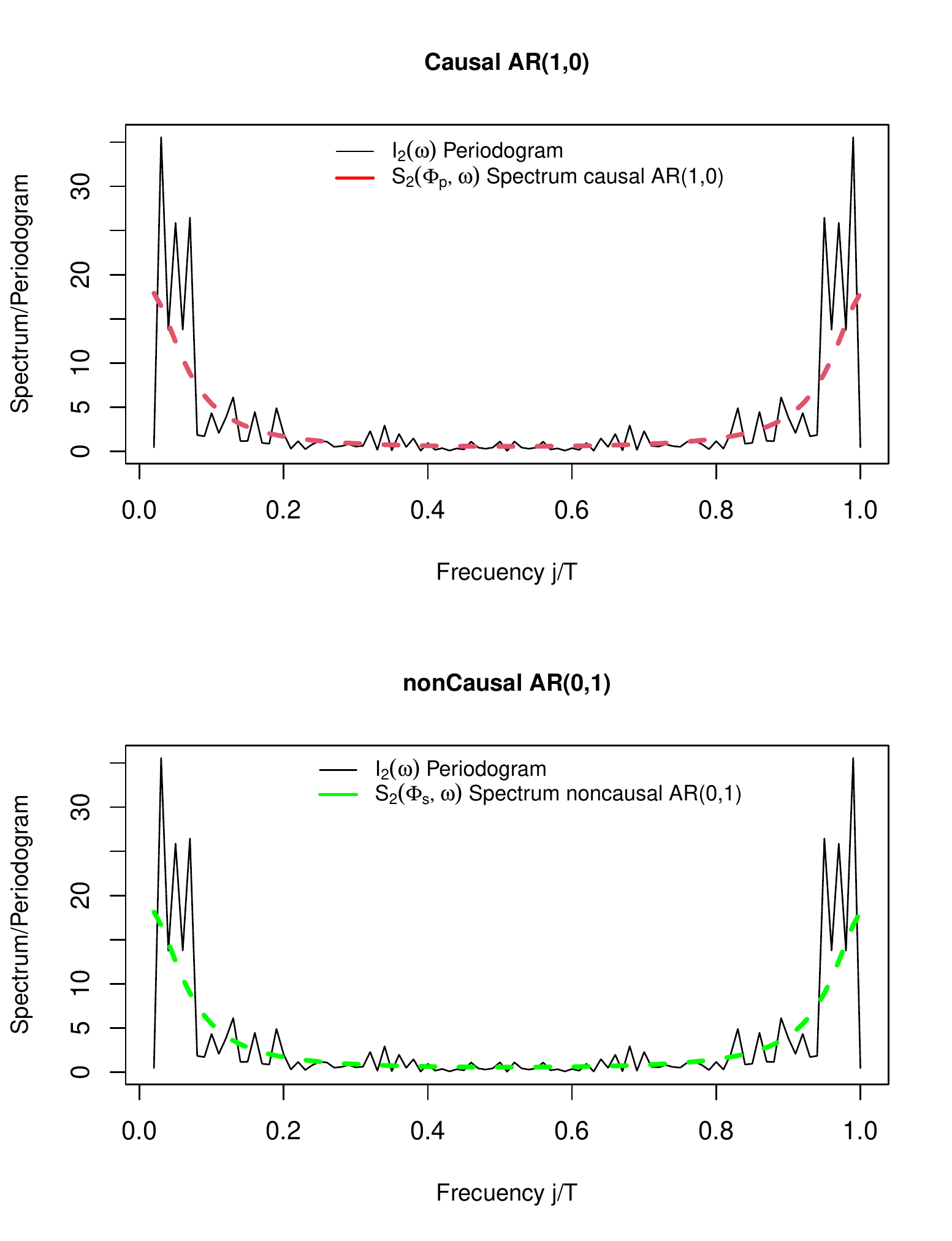}
         \caption{\footnotesize The top Figure is the periodogram and spectrum of a causal AR(1,0) $y_t=0.7y_{t-1}+\varepsilon_t$. The bottom Figure is the periodogram and spectrum of a noncausal AR(0,1) $y_t=0.7y_{t+1}+\epsilon_t$. $\varepsilon_t\sim$ alpha-stable $f_a(t;\alpha=1.5,\beta=0.25,\gamma=1,\delta=0)$, the sample size $T=100$.}
        \label{spectrum_2s}
     \end{figure}
 \end{center}

In Figure \ref{spectrum_2s}, we show the same models as in Figure \ref{causal-noncausal-acf}, but their frequency domain representation. At the top of the graph is the periodogram and spectrum of the causal AR(1,0) $y_t=0.7y_{t-1}+\varepsilon_t$. At the bottom is the periodogram and spectrum of a noncausal AR(0,1) model $y_t=0.7y_{t+1}+\varepsilon_t$. $\varepsilon_t\sim$ alpha-stable $f_a(t;\alpha=1.5,\beta=0.25,\gamma=1,\delta=0)$, the sample size $T=100$. As in Figure \ref{causal-noncausal-acf}, the same error sequence is used to construct the models. The periodogram, $I_2(\omega)$, as well as the spectrums $S_2(\Phi_r\omega)=S_2(\Phi_s\omega)$ are the same in both cases. Consequently, estimations based on second-order moments identify a causal model, even though the DGP is a noncausal model. In both cases, the low frequencies (below 0.2, and above 0.8) contain most of the information in the spectrum. In contrast, the high frequencies do not contain information about the process.

\subsection{Mixed Cuasal/noncausal MAR($r,s$)}
\label{marrs}
Let $y_t$ be a mixed causal-noncausal autoregressive process MAR($r,s$) generated by
\begin{equation}\label{eq11}
    \varphi(L^{-1})\phi(L)y_t=\varepsilon_t,
\end{equation}

where $\phi(L)=1-\phi_1L-\phi_2L^2...\phi_r L^r$ corresponds to the causal component, and $\varphi(L^{-1})=1-\varphi_1L^{-1}-\varphi_2L^{-2}...\varphi_s L^{-s}$ to the noncausal. The traditional AR($p$) model is decomposed to the MAR($r,s$) model by the equivalence of its order $p=r+s$. The set of parameters is $\Upsilon_{r,s}=\left\{\phi_1,...,\phi_r,\varphi_1,...,\varphi_s\right\}$. The MAR($r,s$) is a combination of the variable $y_t$ with $r$ lags and $s$ leads, and both polynomials $\varphi(z)$ and $\phi(z)$, with roots outside the unit circle. Alternatively, the polynomial $\varphi(z)$ has a causal representation with all its roots inside the unit circle, and the polynomial $\phi(z)$ outside the unit circle. For details, see \cite{lii1996maximum}. Accordingly, $y_t$ has a two-sided moving average representation as in Equation \ref{eq2}. The polynomial of the MAR($r,s$) model is $\upsilon_{r,s}(z)=(1-\varphi_1z-\varphi_2z^{2}...-\varphi_sz^{s})(1-\phi_1z-\phi_2z^2...-\phi_r z^r)$, this interaction between the components of the transfer function results in nonlinearities that can hinder the estimation of the parameters. 

The MAR($r,s$) has a causal purely causal representation AR($r+s$,0), as follows

\begin{equation}
\label{caus2}
    \upsilon^{*}_{r+s}(z)y_t=\xi^{*}_t.
\end{equation}

This model representation can be found in \cite{gourieroux2018misspecification}, Appendix A. Where $\xi^{*}_t=-1/\varphi_s\varepsilon_{t-s}$, and $\upsilon^{*}_{r+s}$ is the polynomial of the causal AR($r+s$,0) representation. To obtain it, we follow several steps. First, we apply distributive law for polynomial $\upsilon_{r,s}(L)$

\begin{multline*}
 \upsilon_{r,s}(L) = 1-\varphi_1 L^{-1}-\varphi_1 L^{-2}-...-\varphi_s L^{-s}-\phi_1L+\phi_1\varphi_1+\phi_1\varphi_2L^{-1}+...+\phi_1\varphi_sL^{1-s}-\phi_2L^2+\phi_2\varphi_1L+\phi_2\varphi_2\\
 +...+
 \phi_2\gamma_sL^{2-s}-\phi_rL^r+\phi_r\varphi_1L^{r-1}+\phi_r\varphi_2L^{r-2}+...+\phi_r\varphi_sL^{r-s}.
\end{multline*}

Reverting the process in $s$ time periods, or equivalently multiplying by $L^s$,

\begin{multline*}
\upsilon_{r+s}(L)=-\varphi_s+\phi_1\varphi_sL+\phi_2\varphi_sL^2+...-\varphi_2L^{s-2}-\varphi_1L^{s-1}+\phi_1\varphi_2L^{s-1}+...+L^s+\phi_1\varphi_1L^s+\phi_2\varphi_2L^s+...-\phi_1L^{s+1}\\
+\phi_2\varphi_1L^{s+1}+...-\phi_2L^{s+2}+...+\phi_r\varphi_sL^{r}+...+\phi_r\varphi_2L^{r+s-2}+\phi_r\varphi_1L^{r+s-1}-\varphi_rL^{r+s}.\\
\end{multline*}

Solving for $y_t$, The closed form of the AR($r+s$,0) is

\begin{multline}\label{upsilon}
     y_t=\varphi_1y_{t-1}+\varphi_2y_{t-2}+...-\frac{\varphi_2}{\varphi_s}y_{t-s+2}+\left(\frac{\phi_1\varphi_2-\varphi_1}{\varphi_s}\right)y_{t-s+1}+\left(\frac{1+\phi_1\varphi_1+\phi_2\varphi_2}{\varphi_s}\right)y_{t-s}+\left(\frac{\phi_2\varphi_1-\varphi_1}{\varphi_s}\right)y_{t-s-1}\\
     -\frac{\phi_2}{\varphi_s}y_{t-s-2}+...+\phi_ry_{t-r}+...+\frac{\phi_r\varphi_2}{\varphi_s}y_{t-r-s+2}+\frac{\phi_r\varphi_1}{\varphi_s}y_{t-r-s+1}-\frac{\phi_r}{\varphi_s}y_{t-r-s}-\frac{\xi_{t-s}}{\varphi_s}.
\end{multline}

The polynomial $\upsilon^{*}_{r+s}(z)$ , has the set of parameters $\Upsilon^{*}_{r+s}=\left\{\upsilon^{*}_1,\upsilon^{*}_2,...\upsilon^{*}_s,...\upsilon^{*}_{r+s}\right\}$, then the AR($r+s$,0) can be written as

\begin{multline}\label{eq14}
    y_t=\upsilon^{*}_1 y_{t-1}+\upsilon^{*}_2 y_{t-2}+...+\upsilon^{*}_{s-2}y_{t-s+2}+\upsilon^{*}_{s-1}y_{t-s+1}+\upsilon^{*}_sy_{t-s}+\upsilon^{*}_{s+1}y_{t-s-1}+\upsilon^{*}_{s+2}y_{t-s-2}+...+\upsilon^{*}_{r}y_{t-r}\\
    +...+\upsilon^{*}_{r+s-2}y_{t-r-s+2}+\upsilon^{*}_{r+s-1}y_{t-r-s+1}+\upsilon^{*}_{r+s}y_{t-r-s}+\xi^{*}_t.
\end{multline}

Note that there is a one-to-one relation between Equations \ref{eq11} and \ref{eq14}. Such representation becomes important in identifying the parameters by correctly selecting the initial values in the estimation. To shed light on this relation, we explore the transformation of the MAR($1,1$) into an AR($2,0$) in the \textbf{Remark 2}.\\

\textbf{Remark 2}. The MAR($1,1$) is represented by
\begin{equation}
    (1-\varphi_1L^{-1})(1-\phi_1L)y_t=\varepsilon_t.
\end{equation}

Multiplying by $L$ and solving for $y_t$, we obtain the AR($2,0$) representation

\begin{equation}
    y_t=\left(\frac{1+\phi_1\varphi_1}{\varphi_1}\right)y_{t-1}-\frac{\phi_1}{\varphi_1}y_{t-2}-\frac{\varepsilon_{t-1}}{\varphi_1}.
\end{equation}

The causal representation is 
\begin{equation}
\label{32}
    y_t=\upsilon^{*}_1y_{t-1}+\upsilon^{*}_2 y_{t-2}+\xi^{*}_t,
\end{equation}
where the set of parameters is $\Upsilon^{*}=\left\{\upsilon^{*}_1=\frac{1+\phi_1\varphi_1}{\varphi_1},\upsilon^{*}_2=-\frac{\phi_1}{\varphi_1}\right\}$, and $\xi^{*}_t=-1/\varphi_1\varepsilon_{t-1}$. Estimating the set of parameters $\upsilon^{*}$ is not trivial, given the nonlinearity in the original model. For a similar representation of Equation \ref{32}, see \cite{gourieroux2018misspecification} proposition 3.1.\\

The spectrum of the MAR($r,s$) is 

\begin{equation} \label{eq1}
\begin{split}
S_2(\Upsilon_{p,s},\omega) & = \frac{k_{2_e}}{2\pi}\frac{1}{\lvert \phi(z)\varphi(z^{-1})\rvert^2} \\
 & = \frac{k_{2_e}}{2\pi}\frac{1}{\lvert(1-\phi_1z-\phi_2z^2...-\phi_r z^r)(1-\varphi_1z^{-1}-\varphi_2z^{-2}...-\varphi_sz^{-s}) \rvert^2}.
\end{split}
\end{equation}

The bispectrum is
\begin{equation} \label{bispectrum3}
\begin{split}
  S_3(\Upsilon_{r,s},\omega_1,\omega_2) & = \frac{k_{3_e}}{(2\pi)^2}\psi(\Upsilon_{r,s},\omega_1)\psi(\Upsilon_{r,s},\omega_2)\overline{\psi(\Upsilon_{r,s},\omega_1+\omega_2)}. 
\end{split}
\end{equation}

where $\psi(\Upsilon_{r,s},\omega)=1/\left[(1-\phi_1z-\phi_2z^2...-\phi_r z^r)(1-\varphi_1z^{-1}-\varphi_2z^{-2}...-\varphi_sz^{-s})\right]$, and $\overline{\psi(\Upsilon_{r,s},-\omega_1-\omega_2)}=1/\left[(1-\phi_1\overline{z}-\phi_2\overline{z}^2...-\phi_r \overline{z}^r)(1-\varphi_1\overline{z}^{-1}-\varphi_2\overline{z}^{-2}...-\varphi_s\overline{z}^{-s})\right]$.

\begin{center}
     \begin{figure}[h]
         \centering         \includegraphics[width=0.9 \textwidth]{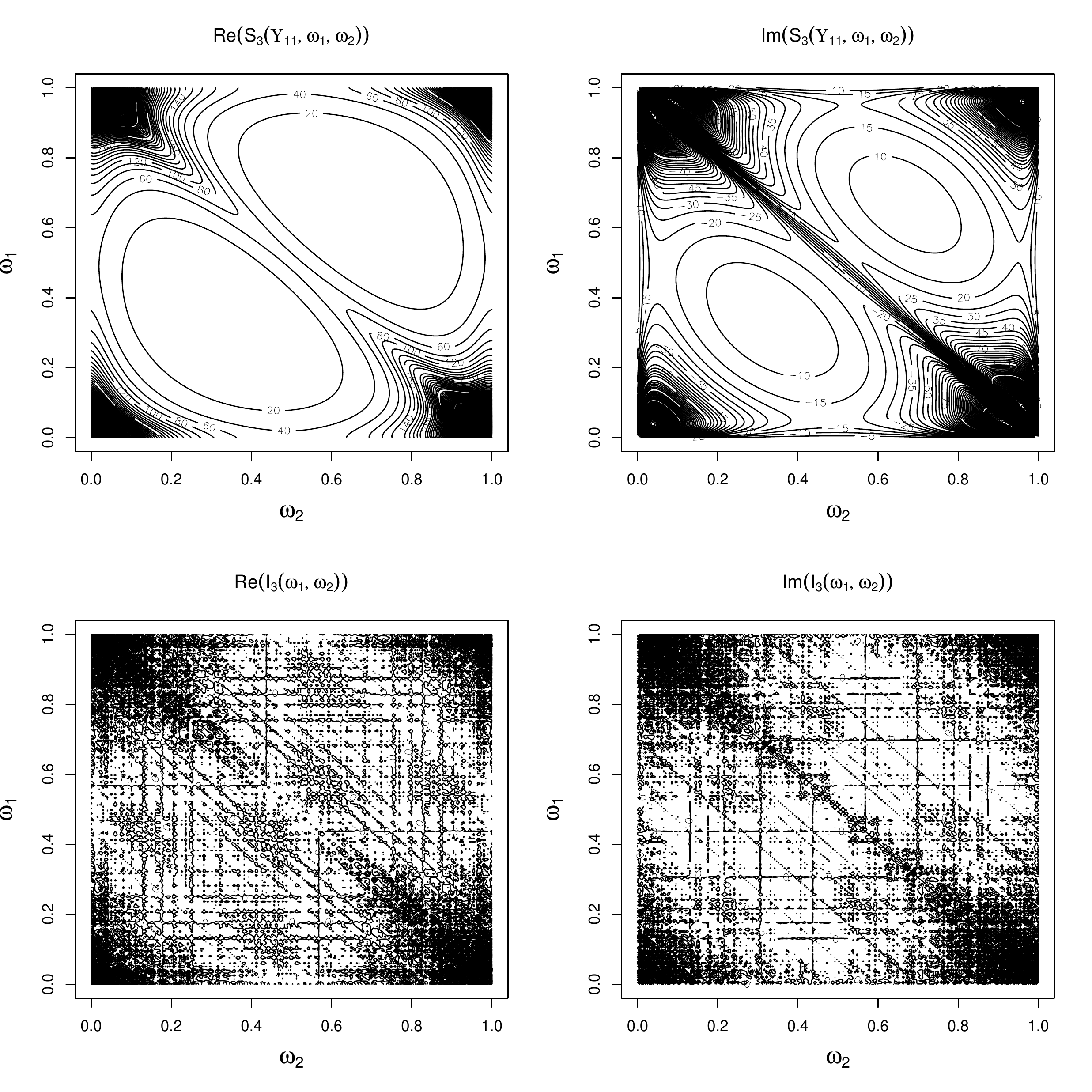}
         \caption{\footnotesize Real and Imaginary part of the bispectrum and biperiodogram for a MAR($1,1$) with $\phi_1=0.7$, $\varphi_1=0.2$, and the error term is alpha-stable $f_a(t;\alpha=1.5,\beta=0.25,\gamma=1,\delta=0)$, the sample size $T=200$.}
        \label{s3_i3}
     \end{figure}
 \end{center}

In Figure \ref{s3_i3}, we show the real and imaginary parts of the bispectrum and the biperiodogram of simulated data from a MAR(1,1) with coefficients $\phi_1=0.7$ and $\varphi_1=0.2$ and $T=200$, the distribution of the errors is alpha-stable $f_a(t;\alpha=1.5,\beta=0.25,\gamma=1,\delta=0)$ with leptokurtic and skewness features. The set of frequencies $\omega_1$ is in the y-axis and $\omega_2$ in the x-axis. In this Figure, the white spaces indicate that the bispectrum value is relatively small and, therefore, the relation between the set of frequencies $\omega_1$ and $\omega_2$ is low. Conversely, when the Figure turns black, there is a higher relation between the set of frequencies. We can observe that the highest amount of information is stored in the low frequencies for $\omega_1$ and $\omega_2$, from 0 to 0.2 and 0.8 to 1 (see the symmetries in Equation \ref{cumu}). This accumulation of information fades as the frequency increases, from 0.2 to 0.5, and from 0.5 to 0.8. In addition, there is an accumulation of information on the $\omega_1=\omega_2$ diagonal that fades as it approaches the frequency 0.5. This same pattern can be seen in the periodogram and spectrum in Figure \ref{spectrum_2s}. Actually, the main diagonal of the biperiodogram (bispectrum) is equivalent to the periodogram (spectrum). Accordingly, the asymmetric patterns that arise from relations between frequencies outside the main diagonal are neglected in the second-order spectrum estimations, thus contributing to the correct estimation of the parameters and the identification of the model in the third-order estimation. Simultaneously, we can observe that the patterns present in the biperiodogram are replicated by the bispectrum. In the real part of the biperiodogram, we can observe two triangular patterns formed by relations between the high frequencies. One below the main diagonal at the frequencies 0.3 to 0.5 and another above the main diagonal at the frequencies 0.5 and 0.7. In addition, four triangular patterns formed a relationship between the high (low) frequencies of $\omega_1$ and low (high) frequencies of $\omega_2$. Two are below the main diagonal, and two are above the main diagonal. Additionally, in the imaginary part of the biperiodogram, there are two triangular patterns at low frequencies, one above and one below the main diagonal. In addition, there are four triangular patterns of the interaction between high frequencies and low frequencies, two above and two below the main diagonal. 

\section{Estimation}
In this Section, we present the estimation function that was first proposed by \cite{leonenko1998spectral}. However, we use the variation presented by \cite{velasco2018frequency} in terms of the scaling factor in the denominators of Equation \ref{est_func}. We define $\boldsymbol{\vartheta}$ as the set of parameters of the autoregressive model, for the causal $\boldsymbol{\vartheta}=\Phi_r$, for the noncausal $\boldsymbol{\vartheta}=\Phi_s$ and for the mixed $\boldsymbol{\vartheta}=\Upsilon_{r,s}$. We first perform a preliminary Gaussian likelihood estimation and obtain $\overline{\boldsymbol{\vartheta}}$, always assuming causality, and we evaluate it in the transfer functions $\psi(\overline{\boldsymbol{\vartheta}},\omega)$ and $\psi(\overline{\boldsymbol{\vartheta}},-\omega_i-\omega_j)$ and the second order cumulant, $\overline{k}_{2}=2\pi T^{-1}\sum_{j=1}^{T-1}I_2(\omega_j)/(\psi(\overline{\boldsymbol{\vartheta}},\omega_j)\psi(\overline{\boldsymbol{\vartheta}},-\omega_j))$. This normalization by preliminary estimates has no effect on the identification of the parameters, given that they are invariant to any inversion of the polynomial roots. This step significantly simplifies the estimation compared with \cite{leonenko1998spectral}. The estimation function is

\begin{equation}
\label{est_func}
R_T(\boldsymbol{\vartheta}) = A_{2T}\sum_{j=1}^{T-1}\left(\frac{ I_2(\omega_j)-S^{*}_2(\boldsymbol{\vartheta},\omega_j)}{ \psi(\overline{\boldsymbol{\vartheta}},\omega_j)\psi(\overline{\boldsymbol{\vartheta}},-\omega_j)}\right)^2+A_{3T}\sum_{j=1}^{T-1}\sum_{i=1}^{T-1}\frac{\lvert I_3(\omega_j,\omega_i)-S^{*}_3(\boldsymbol{\vartheta},\omega_j,\omega_{i}) \rvert^2}{ \psi(\overline{\boldsymbol{\vartheta}},\omega_j)\psi(\overline{\boldsymbol{\vartheta}},\omega_i)\psi(\overline{\boldsymbol{\vartheta}},-\omega_j-\omega_i)},    
\end{equation}

where $A_{2T}=m(2\pi)^2/(4\overline{\kappa_2}^2T)$ and $A_{3T}=n(2\pi)^4/(6\overline{\kappa_2}^3T^2)$. In all cases, $m=1-n$. The weights in our approach are arbitrarily set as $m=0.5$ and $n=0.5$. \cite{terdik1999bilinear} estimate an AR($1,0$) with different values for $m$, highlighting the efficiency gains with the correct selection. Nevertheless, in \cite{velasco2018frequency}, the weights are optimally selected, increasing the efficiency, although they turn out to be unintuitive in some cases, increasing the complexity of the estimation. Considering theorem 2 in \cite{velasco2018frequency}, we include in the spectrum and the bispectrum a consistent estimator of the standardized cumulants of orders two and three, respectively, that is

\begin{equation}
k^{*}_2(\boldsymbol{\vartheta})=\frac{2\pi}{T}\sum_{j=1}^{T-1}\frac{I_2(\omega_j)}{\psi(\boldsymbol{\vartheta},\omega_j)\psi(\boldsymbol{\vartheta},-\omega_j)},\;\; \;\;\;\;k^{*}_3(\boldsymbol{\vartheta})=\frac{4\pi^2}{T^2} \sum_{j=1}^{T-1}\sum_{i=1}^{T-1}Re\left(\frac{I_3(\omega_j,\omega_i)}{\psi(\boldsymbol{\vartheta},\omega_j)\psi(\boldsymbol{\vartheta},\omega_i)\psi(\boldsymbol{\vartheta},-\omega_j-\omega_i)}\right).
\end{equation}

For a correct identification, the parameters and cumulants must be jointly estimated. Therefore, we modify the spectrum function to add the dependence of the cumulants to $\boldsymbol{\vartheta}$. The modified spectrum and bispectrum are

\begin{equation}
S^{*}_2(\boldsymbol{\vartheta},\omega_j)=S_2(\boldsymbol{\vartheta},k^{*}_2(\boldsymbol{\vartheta}),\omega_j),\;\;\;S^{*}_3(\boldsymbol{\vartheta},\omega_j,\omega_i)=S_3(\boldsymbol{\vartheta},k^{*}_3(\boldsymbol{\vartheta}),\omega_j,\omega_i).
\end{equation}

 When the data is Gaussian, the first term of the estimation function $R_T(\boldsymbol{\vartheta})$ (Equation \ref{est_func}) is sufficient to estimate and identify the autoregressive model correctly. The first term in $R_T(\boldsymbol{\vartheta})$ is equivalent to
\cite{whittle1953estimation} or Gaussian-likelihood. Nevertheless, when the data are non-Gaussian, \cite{brillinger1975time} suggests including the second term, that is, the bispectrum estimation. This estimator is consistent and asymptotic normal. For details, see \cite{terdik1999bilinear,velasco2018frequency}. The set of optimal parameters $\hat{\boldsymbol{\vartheta}}$ is obtained by minimizing $R_T$ by the quasi-Newton method

\begin{equation}
\hat{\boldsymbol{\vartheta}}=\argmin_{\boldsymbol{\vartheta} \in \Theta} \: R_T(\boldsymbol{\vartheta}).
    \end{equation}

\subsection{Consistency and asymptotic normality of $R_T(\boldsymbol{\vartheta})$}

For the consistency of the estimator, we use Lemma 1 in \cite{brillinger1974cross}, and Section 4.2, Theorem 73 in \cite{terdik1999bilinear}. Let $\boldsymbol{\vartheta}$ be the true set of parameters, $\boldsymbol{\vartheta} \subset \Theta$, and 

\begin{equation}
R(\boldsymbol{\vartheta})=m\int_{\Lambda_1}\left(\frac{ I_2(\omega_j)-S_2(\boldsymbol{\vartheta},\omega_j)}{ \psi(\overline{\boldsymbol{\vartheta}},\omega_j)\psi(\overline{\boldsymbol{\vartheta}},-\omega_j)}\right)^2d\omega_j+n\int\int_{\Lambda_2}\frac{\lvert I_3(\omega_j,\omega_i)-S_3(\boldsymbol{\vartheta},\omega_j,\omega_{i}) \rvert^2}{ \psi(\overline{\boldsymbol{\vartheta}},\omega_j)\psi(\overline{\boldsymbol{\vartheta}},\omega_i)\psi(\overline{\boldsymbol{\vartheta}},-\omega_j-\omega_i)}d\omega_jd\omega_i.
\end{equation}

\vskip 0.5cm
\textbf{Theorem 1.} \textit{Suppose that $y_t$ has a bispectrum $\lvert S_3(\omega_j,\omega_i)\rvert<\infty$. The first term in $R_T(\boldsymbol{\vartheta})$, $\left(\frac{ I_2(\omega_j)-S^{*}_2(\boldsymbol{\vartheta},\omega_j)}{ \psi(\overline{\boldsymbol{\vartheta}},\omega_j)\psi(\overline{\boldsymbol{\vartheta}},-\omega_j)}\right)^2$, belong to the finite union of closed intervals $\Lambda_1$, and the second term $\frac{\lvert I_3(\omega_j,\omega_i)-S^{*}_3(\boldsymbol{\vartheta},\omega_j,\omega_{i}) \rvert^2}{ \psi(\overline{\boldsymbol{\vartheta}},\omega_j)\psi(\overline{\boldsymbol{\vartheta}},\omega_i)\psi(\overline{\boldsymbol{\vartheta}},-\omega_j-\omega_i)}$  belong to the finite union of compact domains lying inside the open triangle of frequencies $\Lambda_2$. In addition, $R(\boldsymbol{\vartheta})$ has a unique minimum at $\boldsymbol{\vartheta}$. In this circumstances, $R_T(\boldsymbol{\vartheta})\overset{p}{\to}R(\boldsymbol{\vartheta})$ and $\hat{\boldsymbol{\vartheta}}\overset{p}{\to}\boldsymbol{\vartheta}$, as $T\rightarrow\infty$. $\hat{\boldsymbol{\vartheta}}$ is obtained minimizing $R_T(\boldsymbol{\vartheta})$}. 

\vskip 0.5cm

\textit{Under the conditions mentioned above, and assuming that $\hat{\boldsymbol{\vartheta}}$ is close enough to $\boldsymbol{\vartheta}$ when $T\rightarrow\infty$, the Taylor expansion of $R_T(\hat{\boldsymbol{\vartheta}})$ is}

\begin{equation}
    \frac{\partial R_T(\hat{\boldsymbol{\vartheta}})}{\partial \boldsymbol{\vartheta}}=\frac{\partial R_T(\boldsymbol{\vartheta})}{\partial \boldsymbol{\vartheta}}+(\hat{\boldsymbol{\vartheta}}-\boldsymbol{\vartheta})\frac{\partial^2 R_T(\hat{\boldsymbol{\vartheta}})}{\partial^2 \boldsymbol{\vartheta}}.
\end{equation}

\textit{As $\hat{\boldsymbol{\vartheta}}$ minimizes $R_T(\boldsymbol{\vartheta})$, $ \frac{\partial R_T(\hat{\boldsymbol{\vartheta}})}{\partial \boldsymbol{\vartheta}}=0$. Then},
\begin{equation}
    \hat{\boldsymbol{\vartheta}}-\boldsymbol{\vartheta}=-\frac{\frac{\partial R_T(\boldsymbol{\vartheta})}{\partial \boldsymbol{\vartheta}}}{\frac{\partial^2 R_T(\hat{\boldsymbol{\vartheta}})}{\partial^2 \boldsymbol{\vartheta}}}
    \label{consistency}
\end{equation}
\textit{Then}
\begin{equation}
    \hat{\boldsymbol{\vartheta}}=\boldsymbol{\vartheta}+o_P(1)
\end{equation}

The closed forms for the first and second derivatives of Equation \ref{consistency} are \cite{terdik1999bilinear} Section 4. It is reasonable to presume that the numerator in Equation \ref{consistency} is zero since $\hat{\boldsymbol{\vartheta}}$ is close enough to $\boldsymbol{\vartheta}$, as the third-order cumulant is different from zero and the bispectrum exists.

\vskip 0.5cm
\textbf{Theorem 2.} \textit{Under the fulfillment of Theorem 1, $\hat{\boldsymbol{\vartheta}}$ is asymptotically normal as $T\rightarrow\infty$} 

\begin{equation}
    \sqrt{T}(\hat{\boldsymbol{\vartheta}}-\boldsymbol{\vartheta})\overset{D}{\to}N\left(0,\frac{\Sigma_{1}(\boldsymbol{\vartheta})}{\Sigma_{0}(\boldsymbol{\vartheta})}\right),
\end{equation}

where $\Sigma_{0}(\boldsymbol{\vartheta})$ and $\Sigma_{1}(\boldsymbol{\vartheta})$ are Equations 4.17 and 4.18 in \cite{terdik1999bilinear}. For the autoregressive process, the asymptotic variance of $\hat{\boldsymbol{\vartheta}}$ is

\begin{equation}
\begin{split}\label{asymp_var}
   T\textrm{var}(\hat{\boldsymbol{\vartheta}}-\boldsymbol{\vartheta}) & \approx \frac{\left(4m^2+\left[\frac{m^2}{4}\kappa+\frac{n^2}{2}+mn\right]\zeta^2\right)\eta}{\left(2m+\frac{n}{2}\zeta^2\right)^2\eta^2},
\end{split}
\end{equation}

where the skewness ($\zeta$) and kurtosis ($\kappa$) of the error sequence are defined by the standardized third and fourth-order cumulants as
\begin{equation}
\zeta=\frac{\textrm{Cum}(\varepsilon_t,\varepsilon_t,\varepsilon_t)}{(\textrm{var}(\varepsilon_t))^{3/2}},\;\;and\;\;\kappa=\frac{\textrm{Cum}(\varepsilon_t,\varepsilon_t,\varepsilon_t,\varepsilon_t)}{(\textrm{var}(\varepsilon_t))^{2}}.
\end{equation}

The parameter $\eta$ is the Integral respect to $\omega$ of the derivative of the modulus of the second-order transfer function with respect the set of parameters $\boldsymbol{\vartheta}$ 

\begin{equation}
    \eta=\int_{0}^{1/2}\left(\frac{\partial \;}{\partial \boldsymbol{\vartheta}}log\;S_2(\omega,\boldsymbol{\vartheta}) \right)^2d\omega=\int_{0}^{1}\frac{\frac{\partial }{\partial \boldsymbol{\vartheta}}\psi(\boldsymbol{\vartheta},\omega)}{\psi(\boldsymbol{\vartheta},\omega)}\frac{\frac{\partial }{\partial \boldsymbol{\vartheta}}\psi(\boldsymbol{\vartheta},-\omega)}{\psi(\boldsymbol{\vartheta},-\omega)}d\omega.
\end{equation}
 By solving Equation \ref{asymp_var} we can obtain the standard errors. Notice that the asymptotic variance depends on the third and fourth-order cumulants. Since moments and cumulants do not exist for all distributions, we use their sample counterparts when they are absent. \cite{lobato2022single} developed closed formulas for the asymptotic variance of the ARMA model allowing for complex roots. 
\vskip 0.5cm

\subsection{Estimation strategy}\label{est_stra}

Theorem 1 shows that the correct set of parameters can be identified and consistently estimated by finding the global minimum of $R_T(\boldsymbol{\vartheta})$. However, this estimation function is not convex. On the contrary, it is multimodal, leading to the possibility of multiple local minima that increase as the number of parameters in the model grows. As a result, the empirical prerequisite for reaching the global minimum is the correct selection of the initial values, ensuring that $\hat{\boldsymbol{\vartheta}}$ is close enough to the true value $\boldsymbol{\vartheta}$. The latter entails certain limitations for noncausal and mixed models, given their nonlinearity and that $\boldsymbol{\vartheta}$ is not directly observable. Accordingly, the solution of this type of model is not unique and convergent. For a discussion of the multimodality in noncausal models, see \cite{kindop2021ubiquitous}, and for mixed causal-noncausal, see \cite{bec2020mixed}. To illustrate this situation, a model that is originally causal AR($2,0$) has three possible solutions for its parameters. A noncausal AR($0,2$), a mixed MAR($1,1$) and a causal AR($2,0$). Evidently, the solutions AR($0,2$) and MAR($1,1$) correspond to local minima of $R_T(\boldsymbol{\vartheta})$ since the global minimum must be for AR($2,0$). 

The problem of identification in the noncausal model has been studied by \cite{breid1991maximum,andrews2006maximum,lanne2011noncausal} in the time domain. In their procedure, they estimate assuming causality and select the order of the AR($p$). In a second step, they estimate the noncausal AR($0,s$) with $p=s$, and all possible combinations of mixed MAR($r,s$), with $r+s=p$, identifying as the appropriate model the one with the highest log-likelihood. Similarly, we propose an estimation strategy that significantly increases the probability of reaching the global minimum of the estimation function by selecting a set of initial parameters close enough to the true set of parameters, helping $\hat{\boldsymbol{\vartheta}}$ to converge in probability to $\boldsymbol{\vartheta}$. Then, we address the multimodality of the estimation function by selecting the model with the lowest $R_T(\hat{\boldsymbol{\vartheta}})$. The initial parameters are chosen based on the pure causal representation of the noncausal (Equation \ref{caus}) and mixed models (Equation \ref{caus2}) from which their original roots can be reconstructed. In this context, we developed a tool to identify purely causal, noncausal, and mixed autoregressive models in the frequency domain.

For clarity, $R_{T}(\Phi_r)$ refers to the estimation function when the model is purely causal AR($r$,0), $R_{T}(\Phi_s)$ when the model is purely noncausal AR(0,$s$), and $R_T(\Upsilon_{r,s})$ when is mixed causal-noncausal MAR($r,s$). As a common first step, we perform Gaussian likelihood or Whittle estimation and obtain $\overline{\boldsymbol{\vartheta}}$ and $\overline{k}_2$.

\begin{enumerate}
    \item For the estimation of the \textbf{Causal AR($r,0$)}, the selection of the initial values to minimize $R_T(\Phi_r)$ is $\overline{\boldsymbol{\vartheta}}$, that is 

\begin{equation}
\Phi_{p_0}=\overline{\boldsymbol{\vartheta}}=\left\{\phi_{1_0},\phi_{2_0},...,\phi_{p_0}\right\}.
\end{equation}
The initial values are always obtained in the preliminary estimation, assuming causality. The minimum obtained by the estimation function is $R_T(\hat{\Phi}_r)$.

\item For the estimation of the \textbf{noncausal AR($0,s$)}, the selection of the initial values to minimize $R_T(\Phi_s)$ are the coefficients from the polynomial $\varphi^{*}(z)$ in Equation \ref{eq6}, that its causal representation in Equation \ref{caus}

\begin{equation}
    \Phi_{s_0}=\left\{\varphi^{*}_1,\varphi^{*}_2,...,\varphi^{*}_s\right\}.
\end{equation}
We highlight that in \cite{gourieroux2018misspecification}, it is shown that Gaussian-likelihood estimation is a consistent estimator of the causal representation of a noncausal model. After the estimation, the set of parameters $\Phi_s$ is reconstructed with Equation \ref{eq6}. The minimum obtained by the estimation function is $R_T(\hat{\phi}_s)$.

\textbf{Example 1} We want to estimate the noncausal AR($0,2$) $y_t=\varphi_1y_{t+1}+\varphi_2y_{t+2}+\varepsilon_t$, the steps are

\begin{enumerate}
    \item Estimate the causal AR($2,0$) by Gaussian likelihood, obtain $\hat{\varphi}^{*}_1$ and $\hat{\varphi}^{*}_2$.
    
    \item Optimize $R_T(\Phi_s)$ with the initial parameters $\Phi_{s_0}=\left\{\hat{\varphi}^{*}_1,\hat{\varphi}^{*}_2\right\}$.

    \item Recover $\hat{\Phi}_s$ as
    $\hat{\varphi}_1=-\frac{\hat{\varphi}^{*}_1}{\hat{\varphi}^{*}_2}$, $\hat{\varphi}_2=\frac{1}{\hat{\varphi}^{*}_2}$.
\end{enumerate}

\item For the estimation of the \textbf{mixed MAR($r,s$)} the selection of the initial values to minimize $R_T(\Upsilon_{r,s})$ are the set of parameters
\begin{equation}
  \Upsilon_{r_0+s_0}=\left\{\phi_{1_0},...,\phi_{r_0},\varphi_{1_0},...,\varphi_{s_0}\right\}.
\end{equation}
We must first conduct a Gaussian-likelihood to obtain the initial values and obtain $\overline{\boldsymbol{\vartheta}}$. After, the set of parameters $\Upsilon_{r,s}$ is reconstructed using the relation between Equations \ref{upsilon} and \ref{eq14}.

\textbf{Example 2} We want to estimate a mixed causal-noncausal MAR($1,1$), the steps are

\begin{enumerate}
   \item Estimate the causal AR($2,0$) by Gaussian likelihood, and obtain $\hat{\upsilon}^{*}_1$ and $\hat{\upsilon}^{*}_2$.

    \item Recover $\hat{\Upsilon}_{1,1}$ as $\hat{\upsilon}^{*}_1=\frac{1+\phi_1\varphi_1}{\varphi_1}$, and $\hat{\upsilon}^{*}_2=-\frac{\phi_1}{\varphi_1}$. Notice that this nonlinear system has two possible solutions. Both with possible complex roots.

\[\phi_{1_0}=\frac{1}{2}\left(\hat{\upsilon}^{*}_1-\sqrt{(\hat{\upsilon}^{*}_1)^2+4\hat{\upsilon}^{*}_2}\right),\;\textrm{and}\;\;\;\varphi_{1_0}=\frac{-\hat{\upsilon}^{*}_1+\sqrt{(\hat{\upsilon}^{*}_1)^2+4\hat{\upsilon}^{*}_2}}{2\hat{\upsilon}^{*}_2},\]

\[\phi_{1_0}=\frac{1}{2}\left(\hat{\upsilon}^{*}_1+\sqrt{(\hat{\upsilon}^{*}_1)^2+4\hat{\upsilon}^{*}_2}\;\right),\;\textrm{and}\;\;\;\varphi_{1_0}=-\frac{\hat{\upsilon}^{*}_1+\sqrt{(\hat{\upsilon}^{*}_1)^2+4\hat{\upsilon}^{*}_2}}{2\hat{\upsilon}^{*}_2}.\]

\item Optimize $R_T(\Upsilon_{r,s})$ with initial parameters $\Upsilon_{1,1}=\left\{\phi_{1_0},\varphi_{1_0}\right\}$.

\item Obtain $\hat{\Upsilon}_{1,1}=\left\{\hat{\phi}_1,\hat{\varphi}_1\right\}$.

\end{enumerate}
\end{enumerate}

After the estimation, we proceed with the model \textit{\textbf{identification}} based on the value of the estimation function. 

\begin{enumerate}
    \item If the model is causal, then $R_T(\hat{\Phi}_r)=\textrm{min}\left\{R_T(\hat{\Phi}_r),R_T(\hat{\Phi}_s),R_T(\hat{\Upsilon}_{r,s})\right\}$.

    \item If the model is noncausal, then $R_T(\hat{\Phi}_s)=\textrm{min}\left\{R_T(\hat{\Phi}_r),R_T(\hat{\Phi}_s),R_T(\hat{\Upsilon}_{r,s})\right\}$.

    \item If the model is mixed causal-noncausal, then $R_T(\hat{\Upsilon}_{r,s})=\textrm{min}\left\{R_T(\hat{\Phi}_r),R_T(\hat{\Phi}_s),R_T(\hat{\Upsilon}_{r,s})\right\}$.
\end{enumerate}

\section{Simulation study} 
\label{MC}
\begin{center}
     \begin{figure}[h]
         \centering         \includegraphics[width=0.7 \textwidth]{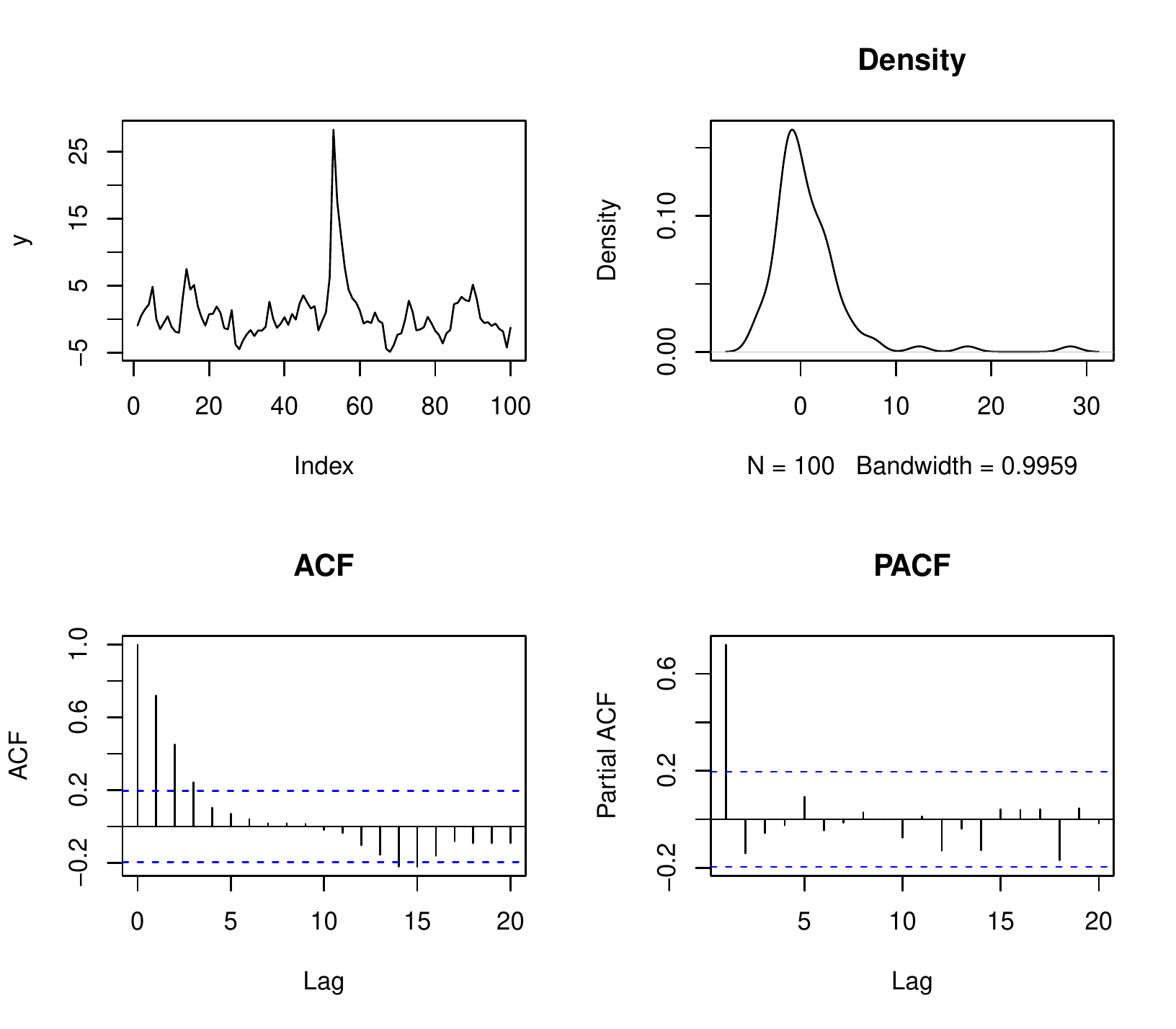}
         \caption{\footnotesize MAR($1,1$) with $\phi_1=0.7$ and $\varphi_1=0.2$, with errors alpha-stable $f_a(t;\alpha=1.5,\beta=0.25,\gamma=1,\delta=0)$, $T=100$.}
        \label{mixed_ma}
     \end{figure}
 \end{center}

We perform an extensive Monte Carlo study to investigate the finite sample properties of our estimation strategy using $R_T(\boldsymbol{\vartheta})$. We consider the ability to identify the three types of models, causal, noncausal and mixed. The results are based on M=1000 simulations. We use the alpha-stable distribution as the data-generating process (DGP). For implementations of the alpha-stable distribution in noncausal models, see \cite{fries2019mixed} and \cite{fries2021conditional}. The probability density function of the alpha-stable is 
\begin{equation}
    f_a(t;\alpha,\beta,\gamma,\delta)=exp(it\delta-\lvert\gamma t\rvert^\alpha(1-i\beta sgn(t)(\lvert\gamma t\rvert^{1-\alpha}-1)tan(\pi\alpha2^{-1}))),
\end{equation}

where $\alpha\in(0,2)$ is the stability parameter, $\beta \left[-1,1\right]$ the skewness parameter, $\gamma>0$ the scale parameter, and $\delta\in\Re $ the location parameter. We generate random numbers with length  $T=\left\{100,200,500\right\}$, for the parameters we choose $\alpha=\left\{1.5,1.8\right\}$, $\beta=0.25$, $\gamma=1$ and $\delta=0$. To allow for skewness and leptokurticity. In Figure \ref{mixed_ma}, we show an alpha-stable MAR($1,1$), the empirical density, the ACF, and the PACF. We can observe
asymmetric cycles, bubble patterns, and high persistence in the ACF. Since we are interested in the ability of our method to identify the correct model, we consider three DGPs, a purely causal AR($2,0$), with $\phi_1=0.7, \phi_2=0.2$, a purely noncausal AR($0,2$), with $\varphi_1=0.7, \varphi_2=0.2$, and a mixed MAR($1,1$) with $\phi_1=0.7, \varphi_1=0.2$.

In Table \ref{MC2}, we present the percentage with which we identify the correct model (identification rate) for the estimation of $R_T(\boldsymbol{\vartheta})$. In the first column, there are the parameters of the alpha-stable distribution, notice that the only parameter that varies is $\alpha$. The second column is the sample size $T$. To compare our method of selecting the initial values and their impact on the identification rate, we compare it with the second-order method, i.e., selecting the initial values directly from the AR($2,0$) Gaussian likelihood. A similar procedure is used in the MARX package \cite{hecq2017simulation}. The identification rate is the number of times that the method succeeds in detecting that the model is causal AR($2,0$) when the DGP is AR($2,0$), given the three possibilities, causal AR($2,0$), noncausal AR($0,2$), and mixed MAR($1,1$). The same logic follows when the DGP is noncausal AR($0,2$) and mixed MAR($1,1$). We report the identification rate for the causal AR($2,0$) in the third column, in the fourth column for the AR($0,2$), and in the fifth column for the MAR($1,1$). 

\begin{table}[h]
\centering
\caption{Rate of Identification of $R_T(\boldsymbol{\vartheta})$}
\label{MC2}
\resizebox{\textwidth}{!}{%
\begin{tabular}{cccccccc}
\hline
                                                            &     & \multicolumn{2}{c}{AR($2,0$)}                                                                                                                    & \multicolumn{2}{c}{MAR($1,1$)}                                                                                                                   & \multicolumn{2}{c}{AR($0,2$)}                                                                                                                    \\ \cline{3-8} 
                                                            &     & \multicolumn{2}{c}{$\phi_1=0.7,\;\phi_2=0.2$}                                                                                                  & \multicolumn{2}{c}{$\phi_1=0.7,\;\varphi_1=0.2$}                                                                                               & \multicolumn{2}{c}{$\varphi_1=0.7,\;\varphi_2=0.2$}                                                                                            \\ \cline{3-8} 
\begin{tabular}[c]{@{}c@{}}alpha-Stable\\ dist\end{tabular} & T   & \begin{tabular}[c]{@{}c@{}}Initial values\\ with our method\end{tabular} & \begin{tabular}[c]{@{}c@{}}Initial values\\ from AR(2)\end{tabular} & \begin{tabular}[c]{@{}c@{}}Initial values\\ with our method\end{tabular} & \begin{tabular}[c]{@{}c@{}}Initial values\\ from AR(2)\end{tabular} & \begin{tabular}[c]{@{}c@{}}Initial values\\ with our method\end{tabular} & \begin{tabular}[c]{@{}c@{}}Initial values\\ from AR(2)\end{tabular} \\ \hline
$\alpha=1.5$                                                & 100 & 74\%                                                                     & 74\%                                                                & 72\%                                                                     & 42\%                                                                & 76\%                                                                     & 10\%                                                                \\
                                                            & 200 & 94\%                                                                     & 94\%                                                                & 80\%                                                                     & 60\%                                                                & 82\%                                                                     & 18\%                                                                \\
                                                            & 500 & 95\%                                                                     & 95\%                                                                & 84\%                                                                     & 66\%                                                                & 86\%                                                                     & 21\%                                                                \\ \hline
$\alpha=1.8$                                                & 100 & 72\%                                                                     & 72\%                                                                & 53\%                                                                     & 39\%                                                                 & 64\%                                                                     & 6\%                                                                \\
                                                            & 200 & 79\%                                                                     & 79\%                                                                & 61\%                                                                     & 44\%                                                                & 72\%                                                                     & 14\%                                                                \\
                                                            & 500 & 88\%                                                                     & 88\%                                                                & 69\%                                                                     & 47\%                                                                & 74\%                                                                     & 18\%                                                                \\ \hline
\end{tabular}%
}
\end{table}

The identification rate reveals that even with 100 observations, our strategy performs well for all three models. In all cases, the percentage increases as the number of data increases. The model with the highest level of identification
is the causal AR($2,0$), followed by the mixed MAR($1,1$) and the noncausal AR($0,2$). As was expected, the
more leptokurtic distribution, $\alpha=1.5$, has a higher identification rate since it deviates more from the Gaussian distribution. The results are equivalent in other simulations with different parameters (not presented). However, the identification rate decays slightly as the parameters approach the unit circle. In the case of the parameters taken from the coefficients of the AR($2,0$) model estimation by Gaussian-likelihood, the identification decreases significantly. In this case, the identification is lower in the noncausal AR($0,2$) model than in the mixed MAR($1,1$) model. It is noteworthy that identification is possible when the sample moments of the data exist, even knowing that for the alpha-stable distribution, only the moments of orders two, three, and four exist when $\alpha=2$. 

The mean and standard deviation of the estimated parameters are reported in Table \ref{MC1}, (we only show the results for our estimation strategy).  

\begin{table}[h]
\caption{Mean and Standard Deviation}
\label{MC1}
\resizebox{\textwidth}{!}{%
\begin{tabular}{llcccccccccccc}
\hline
 &  & \multicolumn{4}{c}{Causal AR(2)} & \multicolumn{4}{c}{noncausal AR(2)} & \multicolumn{4}{c}{MAR($1,1$)} \\ \cline{3-14} 
\multicolumn{1}{c}{\begin{tabular}[c]{@{}c@{}}alpha-Stable\\ dist\end{tabular}} &  & \multicolumn{2}{c}{$\phi_1=0.7$} & \multicolumn{2}{c}{$\phi_2=0.2$} & \multicolumn{2}{c}{$\varphi_1=0.7$} & \multicolumn{2}{c}{$\varphi_2=0.2$} & \multicolumn{2}{c}{$\phi_1=0.7$} & \multicolumn{2}{c}{$\varphi_1=0.2$} \\ \hline
 & \multicolumn{1}{c}{T} & Mean & St.dev & Mean & St.dev & Mean & St.dev & Mean & St.dev & Mean & St.dev & Mean & St.dev \\ \hline
$\alpha=1.5$ & 100 & 0.733 & 0.112 & 0.147 & 0.127 & 0.702 & 0.097 & 0.155 & 0.112 & 0.667 & 0.162 & 0.211 & 0.147 \\
 & 200 & 0.703 & 0.099 & 0.188 & 0.095 & 0.697 & 0.096 & 0.183 & 0.089 & 0.697 & 0.054 & 0.207 & 0.089 \\
 & 500 & 0.700 & 0.055 & 0.196 & 0.058 & 0.700 & 0.062 & 0.195 & 0.060 & 0.701 & 0.031 & 0.200 & 0.042 \\ \hline
$\alpha=1.8$ & 100 & 0.734 & 0.154 & 0.136 & 0.157 & 0.744 & 0.106 & 0.166 & 0.130 & 0.621 & 0.211 & 0.257 & 0.202 \\
 & 200 & 0.727 & 0.091 & 0.170 & 0.110 & 0.731 & 0.090 & 0.157 & 0.091 & 0.699 & 0.077 & 0.191 & 0.105 \\
 & 500 & 0.705 & 0.071 & 0.182 & 0.079 & 0.682 & 0.066 & 0.214 & 0.065 & 0.692 & 0.076 & 0.204 & 0.095 \\ \hline
\end{tabular}%
}
\end{table}

In all three cases, the mean of the estimated parameters is near the true parameters, showing that the estimation is unbiased and revealing the merits of our estimation strategy. The results are better for the more leptokurtic distribution, $\alpha=1.5$, in all cases, meaning that the estimator decreases in bias as the data deviates from normality. As is expected from Theorem 1, the results improve as $T$ increases. In parallel, the standard deviation decreases as $T$ increases. For the causal and noncausal models, the second parameter deviates slightly from its original value when the sample is 100 and 200. The problem is corrected when the sample is 500. In all cases, its standard deviation is lower than the other distribution $\alpha=1.8$, revealing that the variance of the estimator depends on the sample moments. 

\section{Empirical application}

For applying the models discussed above and our estimation strategy, we considered eight monthly commodity prices, including Cotton, measured in US cents per pound. Wheat, Barley, Zinc, Copper, Aluminium, and Nickel, are measured in US dollars per metric ton. We also consider the Brent oil, measured in US dollars per barrel. The eight series are from January 1900 through July 2022 for 383 observations. The data was obtained from the FRED (\url{https://fred.stlouisfed.org/}). 

An accurate understanding of the dynamics of commodity prices is critical to setting the macroeconomic strategies of exporting and importing countries. In many of them, the stability of their national accounts depends on their behavior. The factors that affect commodity prices are the climate, geopolitical events, the internal policies of each country, transport costs, demand, supply, and profitability, among others. Although commodities are a competitive market with many buyers and sellers, there is evidence that their dynamics can be explained with noncausal autoregressive models. \cite{hecq2021forecasting} found evidence of noncausality in monthly Nickel prices, \cite{hecq2019predicting,gourieroux2021convolution} in crude oil monthly prices, \cite{karapanagiotidis2014dynamic} in 25 commodity futures price, including soft, precious metals, energy, and livestock sectors and \cite{lof2017noncausality} in the exchange rates of commodity exporters.

\begin{center}
     \begin{figure}[h]
         \centering         \includegraphics[width=1 \textwidth]{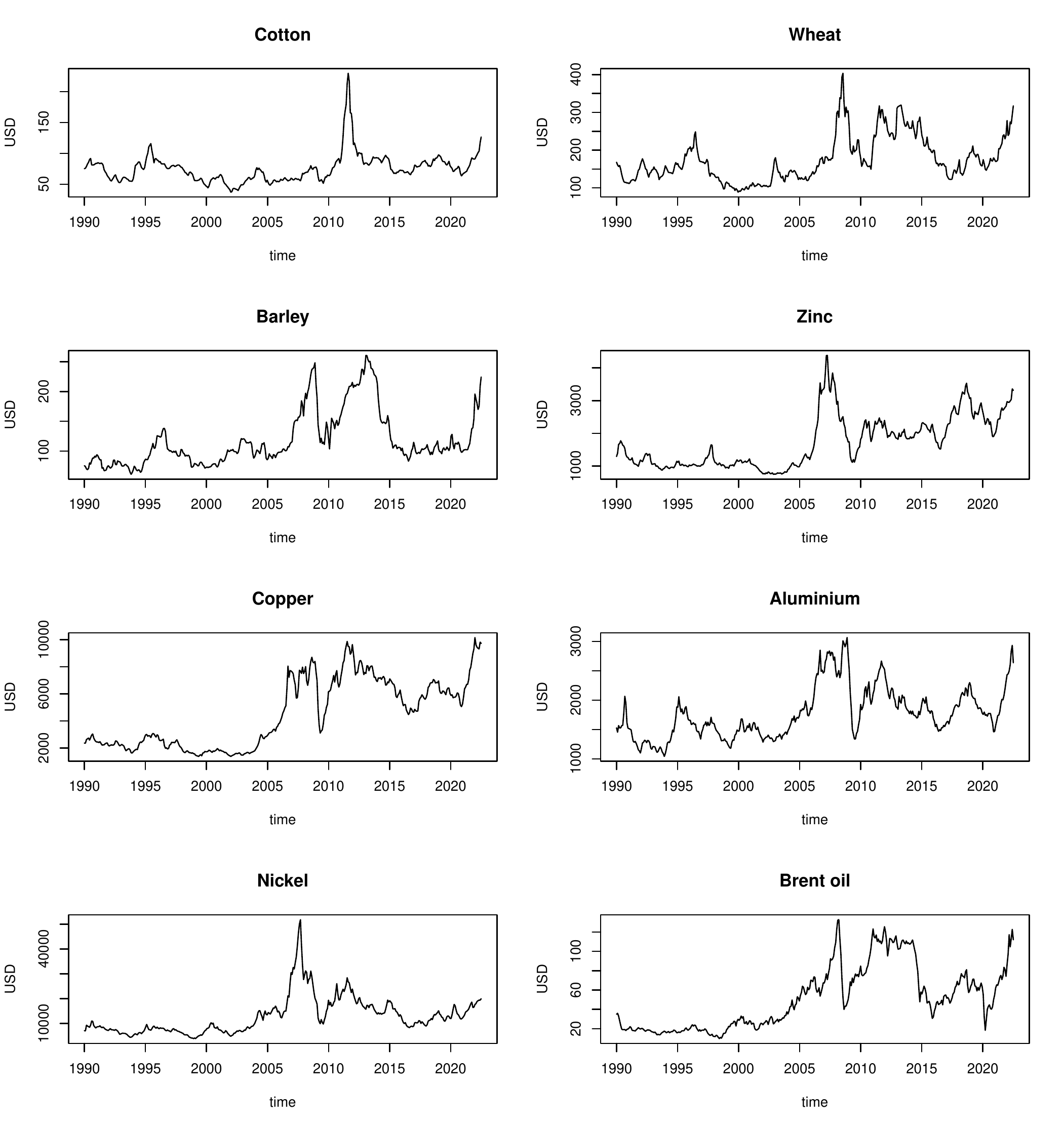}
         \caption{Prices of the eight commodities}
        \label{prices}
     \end{figure}
 \end{center}

All series do not share the same trend, but they appear to be affected by similar shocks. We can observe in Figure \ref{prices} peaks during the 2008 financial crisis, then a drop around 2010 and a subsequent increase from late 2010 to early 2014, then generalized increases in 2022 due to the recent Russian invasion of Ukraine. For the eight series, the null hypothesis of unit root is not 
rejected in the Augmented Dickey-Fuller test with constant and trend alternatives. 

\begin{center}
     \begin{figure}[h]
         \centering         \includegraphics[width=1 \textwidth]{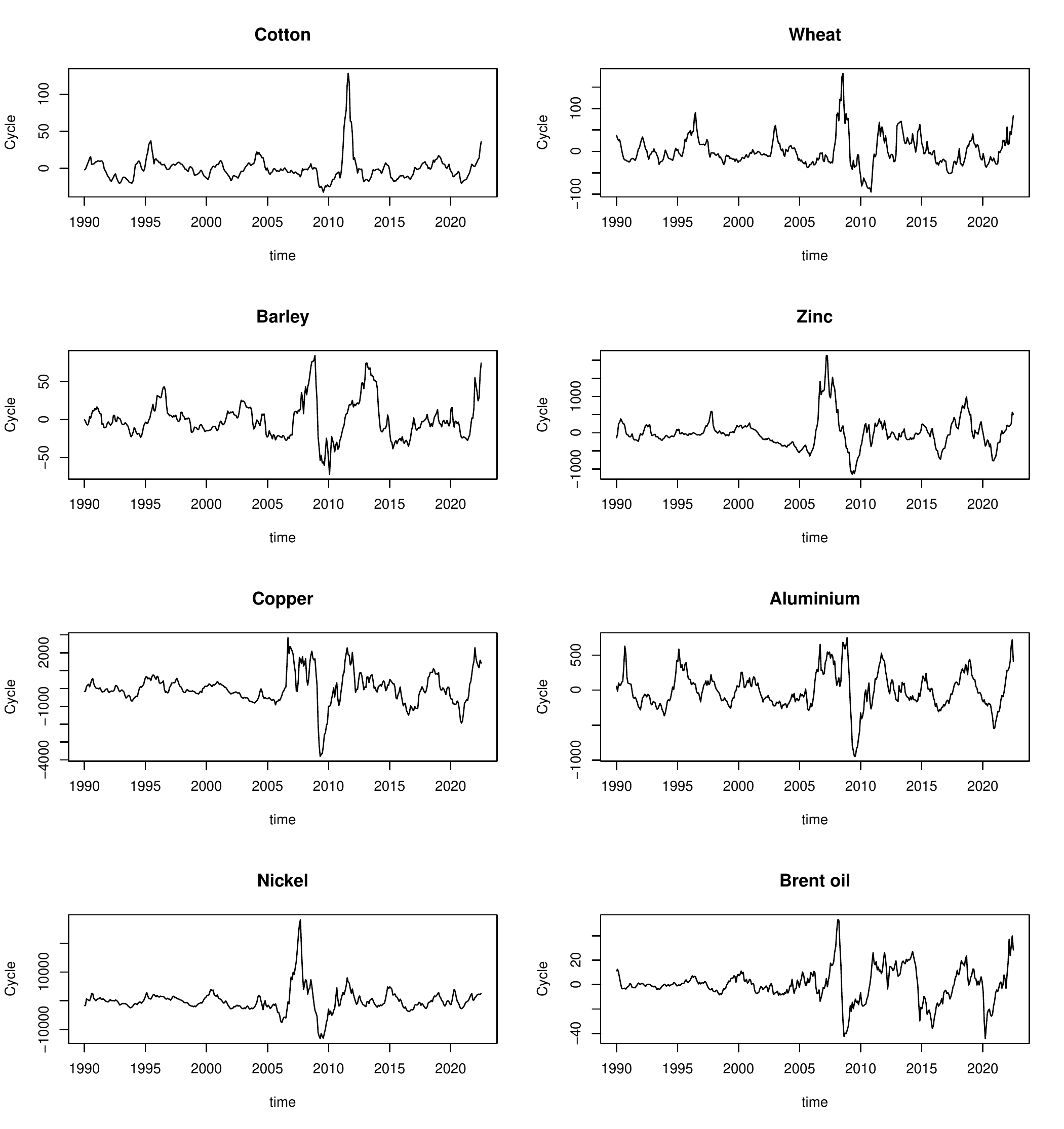}
         \caption{HP cyclical component}
        \label{cycle}
     \end{figure}
 \end{center}

\begin{center}
     \begin{figure}[h]
         \centering         \includegraphics[width=1 \textwidth]{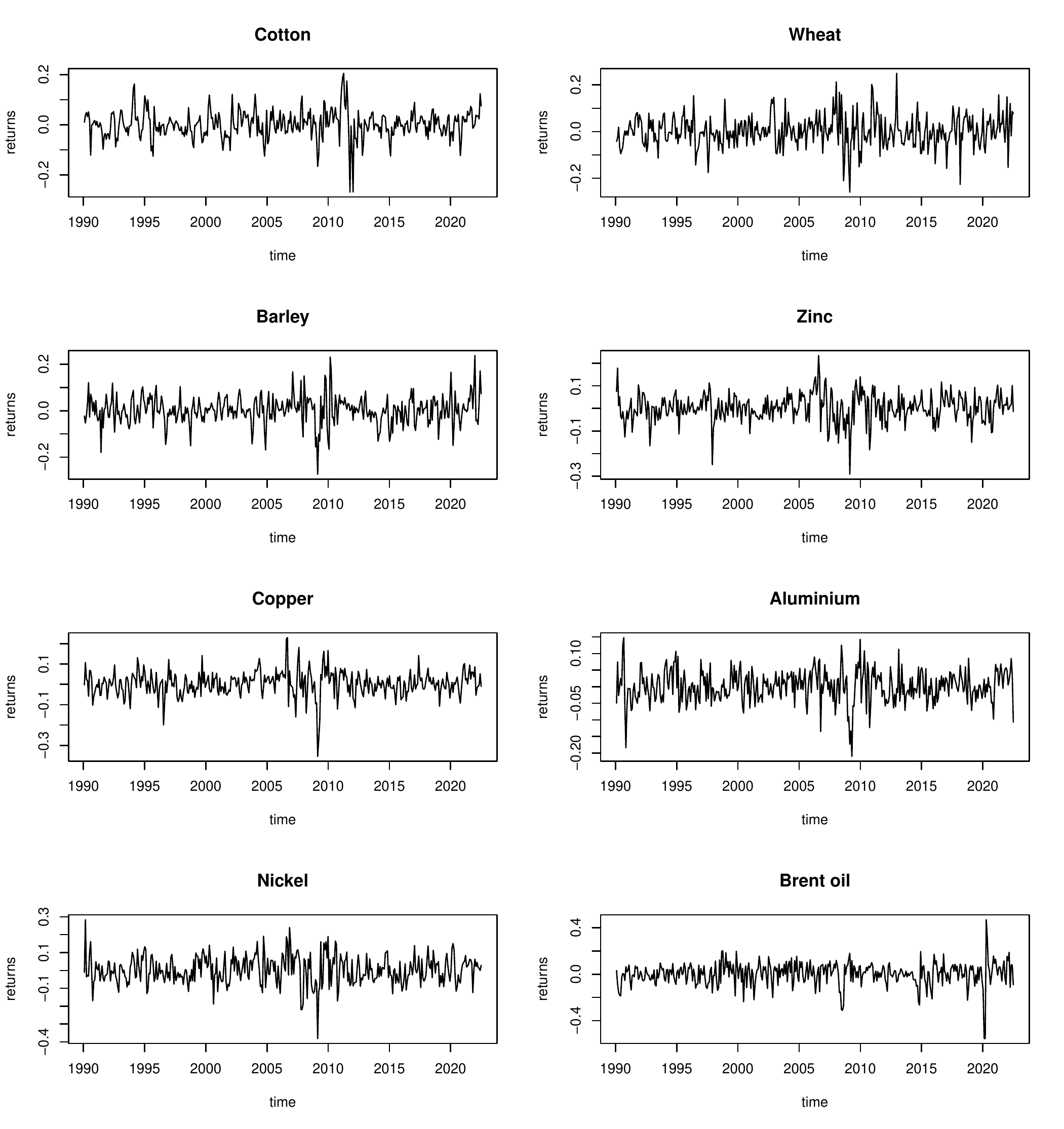}
         \caption{Log-returns}
        \label{returns}
     \end{figure}
 \end{center}

As the data seems non-stationary, a transformation is necessary for the estimation. For instance, \cite{hencic2015noncausal} use a cubic trend, \cite{fries2019mixed} use a linear trend or a simple centering, \cite{hecq2020mixed, cubadda2019detecting} use the returns, \cite{hecq2017seasonal} use the X-11 method to filtering the seasonal effects. However, arbitrarily selecting the type of trend or the transformation of the data can affect the dynamics of the series, greatly influencing the noncausal component. Alternatively, \cite{hecq2019predicting} proposes to use the Hodrick-Prescott (HP hereafter) filter with a penalty parameter $\lambda=$129.600 for monthly data, indicating through a Monte Carlo study that the method keeps the data dynamics unaltered. Notwithstanding the above, there are criticisms about the dynamics induced by the HP filter and the selection of its penalty parameter; see \cite{hamilton2018you} for details.  For the sake of discussion, we consider two transformations, the HP filter with a penalty parameter $\lambda= 129.600$, using the cyclical component, and the logarithmic returns. The Figures of these two transformations are respectively \ref{cycle}, \ref{returns}. 

Following \cite{breid1991maximum}, we estimate all series by Gaussian likelihood and selected the order $p$  by second-order information criteria, AIC and BIC. After selecting the order, we fit the AR($p$) model and obtain the residuals. The descriptive statistics, the selected $p$, and the p-value of the Ljung-Box test for the residuals are in Table \ref{desc}.

\begin{table}[h]
\footnotesize
\centering
\caption{Descriptive statistics and Ljung-Box p-value for the selected AR($p$)}
\label{desc}
\begin{tabular}{lccccccc}
\hline
\multicolumn{1}{c}{} & \multicolumn{7}{c}{Cyclical component HP filter}                                                                                                                \\ \cline{2-8} 
                     & \multicolumn{4}{c}{Descriptive Statistics} & Selected $p$ & \multicolumn{2}{c}{\begin{tabular}[c]{@{}c@{}}Residuals AR($p$)\\ Ljung-Box\\ p-value\end{tabular}} \\ \hline
\multicolumn{1}{c}{} & Mean      & Std.Dev  & Skewness & Kurtosis &              & Lag-1                                            & Lag-2                                            \\ \cline{2-8} 
Cotton               & 7.6E-16   & 17.20    & 3.37     & 21.99    & 2            & 1.000                                            & 0.454                                            \\
Wheat                & -4.3E-15  & 34.72    & 1.17     & 6.95     & 2            & 0.172                                            & 0.225                                            \\
Barley               & 1.4E-15   & 24.94    & 0.83     & 4.33     & 2            & 0.156                                            & 0.224                                            \\
Zinc                 & -2.5E-14  & 423.35   & 1.34     & 7.94     & 2            & 0.344                                            & 0.144                                            \\
Copper               & -1.4E-13  & 889.27   & -0.39    & 6.08     & 2            & 0.273                                            & 0.462                                            \\
Aluminium            & -5.5E-15  & 262.35   & -0.01    & 4.31     & 2            & 0.042                                            & 0.121                                            \\
Nickel               & -3.3E-13  & 4151.15  & 2.13     & 16.77    & 2            & 0.513                                            & 0.733                                            \\
Brent oil            & 1.1E-15   & 13.65    & 0.05     & 4.93     & 2            & 0.660                                            & 0.600                                            \\ \hline
\multicolumn{8}{c}{Log-returns}                                                                                                                                                        \\ \cline{2-8} 
                     & \multicolumn{4}{c}{Descriptive Statistics} & Selected $p$ & \multicolumn{2}{c}{\begin{tabular}[c]{@{}c@{}}Residuals AR($p$)\\ Ljung-Box\\ p-value\end{tabular}} \\ \hline
                     & Mean      & Std.Dev  & Skewness & Kurtosis &              & Lag-1                                            & Lag-2                                            \\ \cline{2-8} 
Cotton               & 0.001     & 0.055    & -0.334   & 6.552    & 1            & 0.262                                            & 0.204                                            \\
Wheat                & 0.002     & 0.067    & 0.065    & 4.514    & 1            & 0.736                                            & 0.630                                            \\
Barley               & 0.003     & 0.061    & -0.113   & 5.371    & 2            & 0.443                                            & 0.324                                            \\
Zinc                 & 0.002     & 0.061    & -0.450   & 5.157    & 2            & 0.549                                            & 0.483                                            \\
Copper               & 0.004     & 0.061    & -0.593   & 7.810    & 1            & 0.953                                            & 0.919                                            \\
Aluminium            & 0.001     & 0.048    & -0.353   & 4.533    & 1            & 0.367                                            & 0.521                                            \\
Nickel               & 0.003     & 0.079    & -0.111   & 4.412    & 2            & 0.853                                            & 0.957                                            \\
Brent oil            & 0.003     & 0.099    & -0.983   & 8.956    & 2            & 0.870                                            & 0.615                                            \\ \hline
\end{tabular}
\end{table}

The series are not non-Gaussian for both transformations. This result was corroborated by the Jarque-Bera test (not presented). In the HP transformation, the highest standard deviations are for Nickel and Copper. All the series have a positive skewness except for Copper and Aluminium. Cotton and Nickel are the series with the highest kurtosis. The selected order is $p=2$ in all cases. We construct the residuals for the AR($2$) model and calculate the Ljung-Box test; the p-value for the first two lags is reported. For all the cases, the null hypothesis is not rejected. 

For the log returns, the standard deviation is relatively similar for all cases, with Brent and Nickel having the highest values. All the series have negative skewness except for Wheat. All series are leptokurtic, with the highest kurtosis for Brent and Copper. In four cases, the selected order $p=1$, and in four cases, $p=2$. The AR($p$) model residuals are calculated, and we report the p-value of the first two lags of the Ljung-Box test. For all the cases, the null hypothesis is not rejected. For the cyclical component of the HP filter, the competing models are the AR($2,0$), the MAR($1,1$), and the AR($0,2$). In Table \ref{hp_est}, we report the estimation using $R_T(\boldsymbol{\vartheta})$ of the three models. The standard deviation of the coefficients is in parentheses. We also present the residual sum of squares (SSE) and the value of the estimation function for each model.

\begin{table}[h]
\caption{Estimation of $R_T(\boldsymbol{\vartheta})$ for the HP cyclical component}
\label{hp_est}
\resizebox{\textwidth}{!}{%
\begin{tabular}{llcccccccc}
\hline
 &  & \multicolumn{8}{c}{Commodity} \\ \cline{3-10} 
Model &  & Cotton & Wheat & Barley & Zinc & Copper & Aluminium & Nickel & Brent oil \\ \hline
AR($2,0$) & $\phi_1$ & \begin{tabular}[c]{@{}c@{}}1.375\\ (0.066)\end{tabular} & \begin{tabular}[c]{@{}c@{}}1.120\\ (0.040)\end{tabular} & \begin{tabular}[c]{@{}c@{}}1.227\\ (0.045)\end{tabular} & \begin{tabular}[c]{@{}c@{}}1.332\\ (0.059)\end{tabular} & \begin{tabular}[c]{@{}c@{}}1.319\\ (0.058)\end{tabular} & \begin{tabular}[c]{@{}c@{}}1.319\\ (0.058)\end{tabular} & \begin{tabular}[c]{@{}c@{}}1.372\\ (0.066)\end{tabular} & \begin{tabular}[c]{@{}c@{}}1.282\\ (0.050)\end{tabular} \\
 & $\phi_2$ & \begin{tabular}[c]{@{}c@{}}-0.463\\ (0.045)\end{tabular} & \begin{tabular}[c]{@{}c@{}}-0.307\\ (0.049)\end{tabular} & \begin{tabular}[c]{@{}c@{}}-0.308\\ (0.049)\end{tabular} & \begin{tabular}[c]{@{}c@{}}-0.404\\ (0.046)\end{tabular} & \begin{tabular}[c]{@{}c@{}}-0.417\\ (0.046)\end{tabular} & \begin{tabular}[c]{@{}c@{}}-0.412\\ (0.047)\end{tabular} & \begin{tabular}[c]{@{}c@{}}-0.457\\ (0.045)\end{tabular} & \begin{tabular}[c]{@{}c@{}}-0.386\\ (0.045)\end{tabular} \\
 &  &  &  &  &  &  &  &  &  \\
 & SSE & 8945.5 & 67963 & 20057 & 6193530 & 32314949 & 2859360 & 596755528 & 8479.296 \\
 & $R_T(\hat{\Phi}_{p})$ & 33.457 & 32.978 & 32.563 & 35.435 & 33.037 & 33.337 & 33.538 & 35.808 \\ \cline{2-10} 
MAR($1,1$) & $\phi_1$ & \begin{tabular}[c]{@{}c@{}}0.496\\ (0.045)\end{tabular} & \begin{tabular}[c]{@{}c@{}}0.423\\ (0.046)\end{tabular} & \begin{tabular}[c]{@{}c@{}}0.319\\ (0.048)\end{tabular} & \begin{tabular}[c]{@{}c@{}}0.484\\ (0.044)\end{tabular} & \begin{tabular}[c]{@{}c@{}}0.472\\ (0.045)\end{tabular} & \begin{tabular}[c]{@{}c@{}}0.511\\ (0.044)\end{tabular} & \begin{tabular}[c]{@{}c@{}}0.564\\ (0.042)\end{tabular} & \begin{tabular}[c]{@{}c@{}}0.459\\ (0.043)\end{tabular} \\
 & $\varphi_1$ & \begin{tabular}[c]{@{}c@{}}0.844\\ (0.027)\end{tabular} & \begin{tabular}[c]{@{}c@{}}0.810\\ (0.030)\end{tabular} & \begin{tabular}[c]{@{}c@{}}0.880\\ (0.024)\end{tabular} & \begin{tabular}[c]{@{}c@{}}0.859\\ (0.026)\end{tabular} & \begin{tabular}[c]{@{}c@{}}0.820\\ (0.029)\end{tabular} & \begin{tabular}[c]{@{}c@{}}0.810\\ (0.030)\end{tabular} & \begin{tabular}[c]{@{}c@{}}0.813\\ (0.030)\end{tabular} & \begin{tabular}[c]{@{}c@{}}0.811\\ (0.029)\end{tabular} \\
 &  &  &  &  &  &  &  &  &  \\
 & SSE & 9022.2 & 68549.9 & 20034 & 6215542 & 32148623 & 2800194 & 596483019 & 8329.35 \\
 & $R_T(\hat{\Upsilon}_{p,s})$ & 32.956 & 32.898 & 32.415 & 35.163 & 32.944 & 33.241 & 33.084 & 35.755 \\ \cline{2-10} 
AR($0,2$) & $\varphi_1$ & \begin{tabular}[c]{@{}c@{}}1.317\\ (0.058)\end{tabular} & \begin{tabular}[c]{@{}c@{}}1.199\\ (0.041)\end{tabular} & \begin{tabular}[c]{@{}c@{}}1.055\\ (0.018)\end{tabular} & \begin{tabular}[c]{@{}c@{}}1.332\\ (0.059)\end{tabular} & \begin{tabular}[c]{@{}c@{}}1.312\\ (0.057)\end{tabular} & \begin{tabular}[c]{@{}c@{}}1.315\\ (0.057)\end{tabular} & \begin{tabular}[c]{@{}c@{}}1.378\\ (0.067)\end{tabular} & \begin{tabular}[c]{@{}c@{}}1.293\\ (0.052)\end{tabular} \\
 & $\varphi_2$ & \begin{tabular}[c]{@{}c@{}}-0.416\\ (0.047)\end{tabular} & \begin{tabular}[c]{@{}c@{}}-0.308\\ (0.049)\end{tabular} & \begin{tabular}[c]{@{}c@{}}-0.143\\ (0.051)\end{tabular} & \begin{tabular}[c]{@{}c@{}}-0.404\\ (0.046)\end{tabular} & \begin{tabular}[c]{@{}c@{}}-0.409\\ (0.047)\end{tabular} & \begin{tabular}[c]{@{}c@{}}-0.408\\ (0.047)\end{tabular} & \begin{tabular}[c]{@{}c@{}}-0.463\\ (0.045)\end{tabular} & \begin{tabular}[c]{@{}c@{}}-0.398\\ (0.045)\end{tabular} \\
 &  &  &  &  &  &  &  &  &  \\
 & SSE & 9078.3 & 67759.8 & 20943 & 5823797 & 32116060 & 2790389 & 596800743 & 8340.1 \\
 & $R_T(\hat{\Phi}_s)$ & 31.720 & 32.984 & 32.474 & 35.381 & 33.054 & 33.331 & 33.391 & 35.795 \\ \hline
\multicolumn{1}{c}{\begin{tabular}[c]{@{}c@{}}Selected\\ model\end{tabular}} & \multicolumn{1}{c}{} & AR($0,2$) & MAR($1,1$) & MAR($1,1$) & MAR($1,1$) & MAR($1,1$) & MAR($1,1$) & MAR($1,1$) & MAR($1,1$) \\ \hline

\multicolumn{1}{c}{\begin{tabular}[c]{@{}c@{}}Selected model\\by MLE with MARX\end{tabular}} & \multicolumn{1}{c}{} & AR($0,2$) & AR($2,0$) & MAR($1,1$) & MAR($1,1$) & MAR($1,1$) & AR($0,2$) & MAR($1,1$) & AR($0,2$) \\ \hline
\end{tabular}%
}
\end{table}

The results from Table \ref{hp_est} reveal that the model selected in seven cases is MAR($1,1$), except for Cotton, which is a purely noncausal AR($0,2$). A feature of the causal AR($2,0$) estimation is that the coefficient of the first lag is outside the unit circle, and the second lag is negative and inside the unit circle. This result is equivalent in the noncausal model AR($0,2$), with the coefficient of the first lead outside the unit circle and the coefficient of the second lead, negative and inside the unit circle. Notice that the coefficient outside the unit circle does not violate the stationarity assumption since, in all cases, the modulus of the characteristic roots is greater than one. Gaussian-likelihood also obtains this estimation pattern. Alternatively, for the MAR($1,1$), the noncausal coefficient is greater than the causal coefficient, both positive and significant. The causal coefficient is between 0.319 for Barley and 0.564 for Nickel. The noncausal coefficient is more homogeneous and always positive, between 0.810 for Aluminium and 0.880 for Barley. In most cases, the AR($2,0$) model has the lower SSE. In the last row, we add the estimation using the MARX package \cite{hecq2017simulation}, that is, Student's t MLE. The results are equivalent in 5 cases. However, different for Wheat AR($2,0$), Aluminium AR($0,2$), and Brent oil AR($0,2$). Similar identification and coefficients were obtained following the time domain estimation of \cite{lanne2011noncausal}. In Table \ref{logret}, we present the estimation using $R_T(\boldsymbol{\vartheta})$ for the log-returns.

\begin{table}[H]
\caption{Estimation Log-returns}
\label{logret}
\resizebox{\textwidth}{!}{%
\begin{tabular}{llcccccccc}
\hline
 &  & \multicolumn{8}{c}{Commodity} \\ \cline{3-10} 
Model &  & Cotton & Wheat & Barley & Zinc & Copper & Aluminium & Nickel & Brent oil \\ \hline
AR($p$,0) & $\phi_1$ & \begin{tabular}[c]{@{}c@{}}0.415\\ (0.084)\end{tabular} & \begin{tabular}[c]{@{}c@{}}0.207\\ (0.062)\end{tabular} & \begin{tabular}[c]{@{}c@{}}0.370\\ (0.057)\end{tabular} & \begin{tabular}[c]{@{}c@{}}0.268\\ (0.016)\end{tabular} & \begin{tabular}[c]{@{}c@{}}0.401\\ (0.087)\end{tabular} & \begin{tabular}[c]{@{}c@{}}0.332\\ (0.089)\end{tabular} & \begin{tabular}[c]{@{}c@{}}0.327\\ (0.043)\end{tabular} & \begin{tabular}[c]{@{}c@{}}0.308\\ (0.048)\end{tabular} \\
 & $\phi_2$ &  &  & \begin{tabular}[c]{@{}c@{}}-0.184\\ (0.060)\end{tabular} & \begin{tabular}[c]{@{}c@{}}0.041\\ (0.016)\end{tabular} &  &  & \begin{tabular}[c]{@{}c@{}}-0.031\\ (0.045)\end{tabular} & \begin{tabular}[c]{@{}c@{}}-0.193\\ (0.050)\end{tabular} \\
 &  &  &  &  &  &  &  &  &  \\
 & SSE & 0.920 & 1.616 & 1.230 & 1.320 & 1.192 & 0.825 & 2.028 & 3.852 \\
 & $R_T(\hat{\Phi}_{p})$ & 31.924 & 31.741 & 32.166 & 32.529 & 31.413 & 31.528 & 32.076 & 35.574 \\ \cline{2-10} 
MAR($1,1$) & $\phi_1$ &  &  & \begin{tabular}[c]{@{}c@{}}-0.085\\ (0.051)\end{tabular} & \begin{tabular}[c]{@{}c@{}}0.333\\ (0.019)\end{tabular} &  &  & \begin{tabular}[c]{@{}c@{}}0.293\\ (0.054)\end{tabular} & \begin{tabular}[c]{@{}c@{}}-0.146\\ (0.010)\end{tabular} \\
 & $\varphi_1$ &  &  & \begin{tabular}[c]{@{}c@{}}0.352\\ (0.048)\end{tabular} & \begin{tabular}[c]{@{}c@{}}-0.090\\ (0.054)\end{tabular} &  &  & \begin{tabular}[c]{@{}c@{}}0.021\\ (0.057)\end{tabular} & \begin{tabular}[c]{@{}c@{}}0.440\\ (0.028)\end{tabular} \\
 &  &  &  &  &  &  &  &  &  \\
 & SSE &  &  & 1.265 & 1.291 &  &  & 2.031 & 3.986 \\
 & $R_T(\hat{\Upsilon}_{p,s})$ &  &  & 32.169 & 32.554 &  &  & 32.077 & 35.479 \\ \cline{2-10} 
AR(0,$s$) & $\varphi_1$ & \begin{tabular}[c]{@{}c@{}}0.378\\ (0.111)\end{tabular} & \begin{tabular}[c]{@{}c@{}}0.224\\ (0.073)\end{tabular} & \begin{tabular}[c]{@{}c@{}}0.378\\ (0.033)\end{tabular} & \begin{tabular}[c]{@{}c@{}}0.164\\ (0.010)\end{tabular} & \begin{tabular}[c]{@{}c@{}}0.383\\ (0.066)\end{tabular} & \begin{tabular}[c]{@{}c@{}}0.279\\ (0.066)\end{tabular} & \begin{tabular}[c]{@{}c@{}}0.321\\ (0.052)\end{tabular} & \begin{tabular}[c]{@{}c@{}}0.474\\ (0.022)\end{tabular} \\
 & $\varphi_2$ &  &  & \begin{tabular}[c]{@{}c@{}}-0.144\\ (0.035)\end{tabular} & \begin{tabular}[c]{@{}c@{}}0.073\\ (0.011)\end{tabular} &  &  & -0.032 & \begin{tabular}[c]{@{}c@{}}-0.113\\ (0.061)\end{tabular} \\
 &  &  &  &  &  &  &  &  &  \\
 & SSE & 0.920 & 1.612 & 1.225 & 1.309 & 1.192 & 0.814 & 2.030 & 3.937 \\
 & $R_T(\hat{\Phi}_s)$ & 31.870 & 31.739 & 32.162 & 32.527 & 31.430 & 31.539 & 32.077 & 35.486 \\ \hline
\multicolumn{1}{c}{\begin{tabular}[c]{@{}c@{}}Selected\\ model\end{tabular}} & \multicolumn{1}{c}{} & AR(0,1) & AR(0,1) & AR($0,2$) & MAR($1,1$) & AR(1,0) & AR(1,0) & MAR($1,1$) & MAR($1,1$) \\ \hline
\multicolumn{1}{c}{\begin{tabular}[c]{@{}c@{}}Selected model\\by MLE with MARX\end{tabular}} & \multicolumn{1}{c}{} & AR(0,1) & AR(0,1) & AR($0,2$) & MAR($1,1$) & AR(1,0) & AR(0,1) & AR($2,0$) & AR($0,2$) \\ \hline

\end{tabular}%
}
\end{table}

 We identify an AR(1,0) for Cotton and Wheat, AR(0,1) for Copper and Aluminium, one AR($0,2$) for Barley, and three MAR($1,1$) for Zinc, Nickel, and Brent oil. In the case of the AR(0,1) model, the coefficients are 0.415 for Cotton and 0.207 for Wheat. Both are highly significant. Similar results are obtained for the AR(1,0) model, where the coefficients are 0.383 for Copper and 0.279 for Aluminium. For MAR($1,1$), the causal coefficients are 0.333 and 0.293, and -0.146 for Zinc, Nickel, and Brent oil, respectively. For the same commodities, the noncausal coefficients are -0.090, 0.021, and 0.440, respectively. In the last row, we add the identification of the MARX package using the 's t MLE. The results are equivalent in six cases. they are different for Aluminium AR(0,1) and Brent oil AR($0,2$).

It is noted that the possible dynamics found in the data are dependent on the transformation performed. In the case of the HP filter, in seven of the eight possible cases, the model identified in its cyclical component is MAR($1,1$). On the other hand, the log-return transformation results are more heterogeneous, with three MAR($1,1$), two AR($0,1$), one AR($0,2$), and one AR($1,0$). The results are similar using Student's t MLE using the MARX package. In the case of the HP filter, the identification is equivalent in five cases. In the case of the log returns, it is equivalent in six cases.

\begin{center}
     \begin{figure}[h]
         \centering         \includegraphics[width=1 \textwidth]{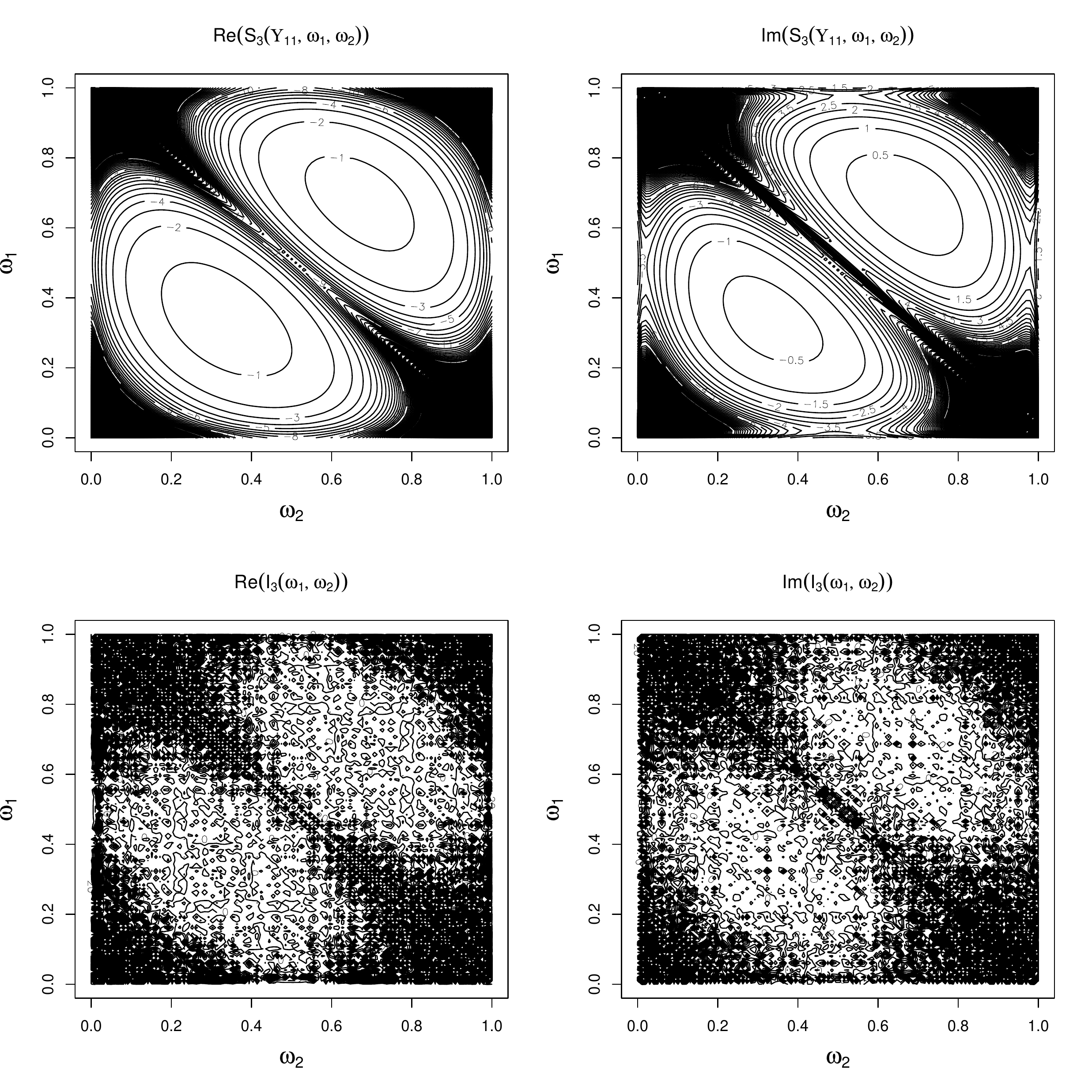}
         \caption{\footnotesize MAR($1,1$) Real and imaginary part of the bispectrum and biperiodogram for the cyclical component of the Brent oil}
        \label{re_bispec}
     \end{figure}
 \end{center}

The real and imaginary parts of the bispectrum and the biperiodogram of the MAR($1,1$) for the Brent oil cyclical component are presented in Figure \ref{re_bispec}. We observed that the highest value relations are found between the low frequencies, 0 to 0.25 and 0.8 to 1. In this range, we observe asymmetric relations, reflecting the non-Gaussianity of the data. No clear relations are observed between the middle frequencies, from 0.25 to 0.4, and the low frequencies. Neither between the high frequencies, from 0.4 to 0.5, with the low frequencies. The same pattern can be observed in the main diagonal of the real and imaginary parts of the biperiodogram. 

\begin{center}
     \begin{figure}[h]
         \centering         \includegraphics[width=0.7 \textwidth]{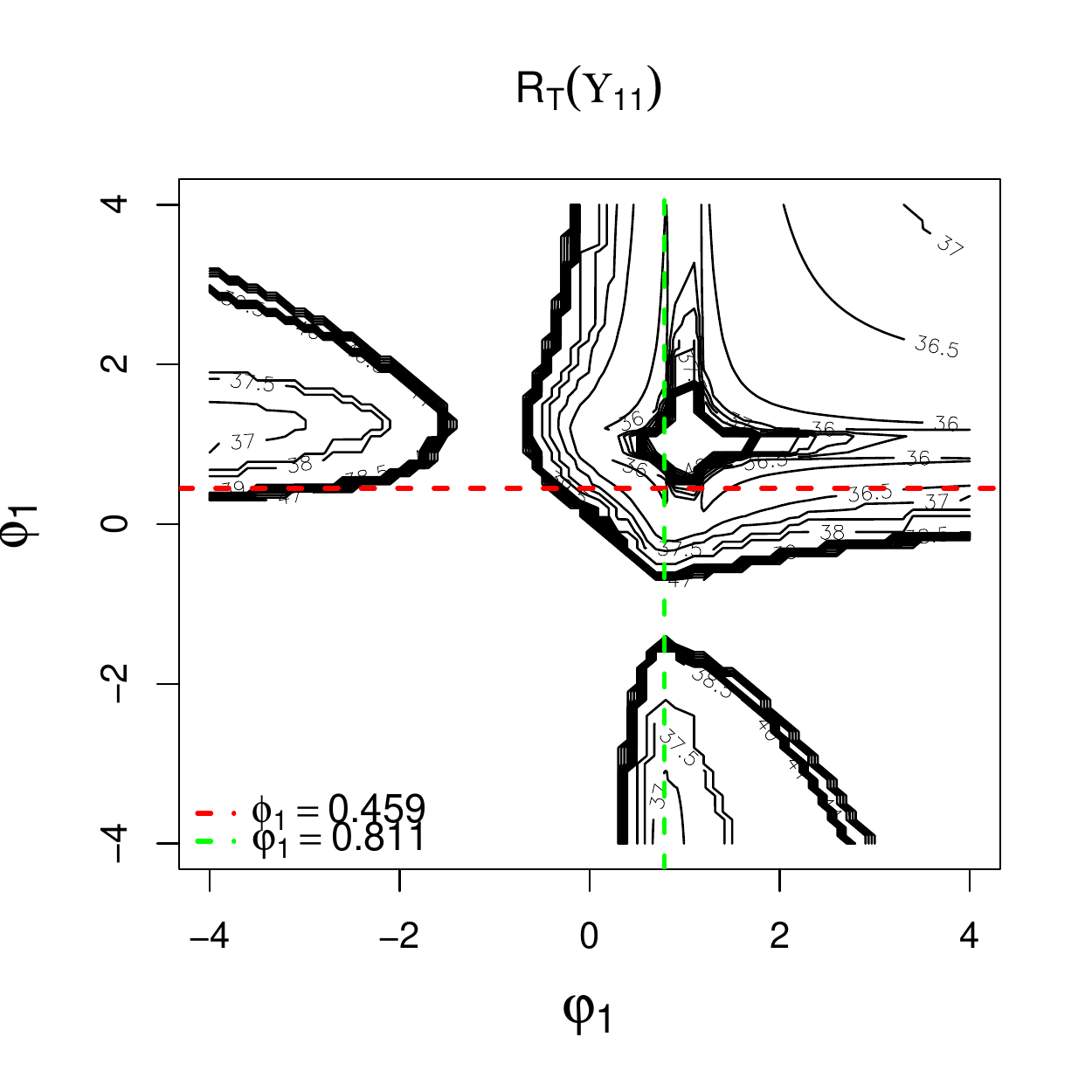}
         \caption{Brent oil MAR($1,1$) cyclical component, estimation function $R_T(\hat{\Upsilon}_{r,s})$}
        \label{rt_upsilon}
     \end{figure}
 \end{center}

In Figure \ref{rt_upsilon}, we present the graphical representation of the estimation function $R_T(\hat{\Upsilon}_{p,s})$ for the cyclical
component of the Brent oil. On the x-axis is the noncausal coefficient ($\varphi_1$), and on the y-axis is the causal coefficient ($\phi_1$). The surface features have been truncated to accentuate the areas where the minima are located. We observe three regions where minima occur. Two of them correspond to local minima, one for the causal model AR(2,0) and one for the noncausal AR(0,2). The global minimum of the estimation function is for the MAR(1,1) with parameters $\varphi_1=0.811$ and $\phi_1=0.459$, located at the intersection of the vertical (green) and horizontal (red) lines.
 
\section{Conclusions and Discussion}
This paper presents a frequency domain estimation method based on a combination of the spectrum and the bispectrum for causal, noncausal, and mixed autoregressive models. We test its properties in finite samples by conducting a Monte Carlo study, finding unbiased estimated parameters and asymptotic normality in the variance. We also performed an empirical application on eight monthly commodity prices finding evidence of noncausality and mixed causality/noncausality. 

Our contribution to the current literature is divided into three parts. First, we develop a general scheme for selecting the initial values for the estimation of noncausal and mixed causal-noncausal based on their causal representations. In our second contribution, we present a decision method for identifying the model based on the existence of the global minimum in the estimation function. Third, we propose the third-order spectral estimation of mixed models.

Our estimation method is consistent with strong asymptotic properties without assuming any probability density function. Instead, non-Gaussianity is required. Nonetheless, it may have a higher computational burden than parametric methods, which assume a distribution function to estimate by ML. Although our method is helpful for estimating causal, noncausal, and mixed models, it depends on the initial values obtained in second-order estimates.

The dynamics of commodities depend on their past history but also show a dependence on future information. Nevertheless, the possible dynamics depend on the transformation performed. In the case of the HP filter, the model identified is a MAR($1,1$) in seven cases. On the other hand, in the log-return transformation, the results are more heterogeneous. It identifies an AR($1,0$) twice, AR($0,1$) twice, one AR($0,2$) and three MAR($1,1$). The dynamics are stronger in the HP filter.

Developing a tool to detect dynamics considering higher-order cumulants is interesting for future research. This is because although the estimation is of third-order, second-order criteria are used to detect dynamics, leading to a process that may not reveal all the information in the data. Also consider the existence of the noninvertible component of the moving average model.

\bibliographystyle{johd}
\bibliography{bib}

\section*{Appendix 1. Figures}

\end{document}